\documentclass[a4paper]{jpconf}
\usepackage{graphicx}
\usepackage[english]{babel}
\usepackage{indentfirst}
\usepackage{afterpage}

\def\Oo {\displaystyle}
\def\gee{ \, \lower 1mm\hbox{$\,{\buildrel > \over{\scriptstyle\scriptstyle\sim} }\displaystyle \,$}}
\def\lee{ \, \lower 1mm\hbox{$\,{\buildrel < \over{\scriptstyle\scriptstyle\sim} }\displaystyle \,$}}
\def\ltsima{$\; \buildrel < \over \sim \;$}
\def\simlt{\lower.5ex\hbox{\ltsima}}
\def\gtsima{$\; \buildrel > \over \sim \;$}
\def\simgt{\lower.5ex\hbox{\gtsima}}

\newcommand{\E}{\mathcal{E}}

\begin{document}
\title{Numerical code for multi-component galaxies: from~$N$-body to chemistry and magnetic fields}

\author{S.A.~Khoperskov $^{1,2}$, E.O.~Vasiliev$^{3}$, A.V.~Khoperskov$^{4}$, V.N.~Lubimov$^{1}$}
\address{Institute of Astronomy of the Russian Academy of Sciences, Pyatnitskaya st., 48, 119017, Moscow, Russia$^{1}$ \\
Sternberg Astronomical Institute, Moscow M.V. Lomonosov State University, Universitetskij pr., 13, Moscow, 119992, Russia$^{2}$ \\
Institute of Physics, Department of Physics, Southern Federal University, Stachki Ave. 194, 344090 Rostov-on-Don, Russia$^{3}$ \\
Volgograd State University, Universitetsky pr., 100, Volgograd 400062, Russia$^{4}$}

\ead{khoperskov@inasan.ru}

\begin{abstract}
We present a numerical code for multi-component simulation of the galactic evolution. Our code includes the following
parts: $N$-body is used to evolve dark matter, stellar dynamics and dust grains, gas dynamics is based on TVD-MUSCL 
scheme with the extra modules for thermal processes, star formation, magnetic fields, chemical kinetics and multi-species
advection. We describe our code in brief, but we give more details for the magneto-gas dynamics. We present several
tests for our code and show that our code have passed the tests with a reasonable accuracy. Our code is parallelized
using the MPI library. We apply our code to study the large scale dynamics of galactic discs. 
\end{abstract}

\section{Introduction}

An explosive growth of observational data on dynamics and chemical composition of various components (subsystems)
of galaxies is seen in last decade. To interpret the huge data cubes we need reasonable models included more
physical processes like magnetic fields, star formation, feedback, chemical kinetics, dust grain dynamics and so on.
Spatial scales of the processes in a galaxy vary from sub-parsecs for turbulent structures in the warm interstellar medium 
to several kiloparsec size for galactic disc and winds. Formation of molecules and distribution of dust grains strongly
affect on the efficiency of star formation, which, in turn, changes the thermal state of a gas through stellar winds and
supernova explosions~\cite{2012ARA&A..50..531K}. Then, to study the interaction between various components numerically someone needs a code with a 
lot of physical modules.

In recent years several open source codes were developed for astrophysical applications, such as ZEUS\footnote{http://www.astro.princeton.edu/~jstone/zeus.html}, 
RAMSES\footnote{http://irfu.cea.fr/Phocea/Vie\_des\_labos/Ast/ast\_sstechnique.php?id\_ast=904}, 
GADGET\footnote{http://www.mpa-garching.mpg.de/gadget/}, Athena\footnote{https://trac.princeton.edu/Athena/}, FLASH\footnote{http://flash.uchicago.edu/site/} and many others. These codes are general-purpose ones, so that each of them has own
advantages and disadvantages in the application to the galactic evolution~\cite{2007MNRAS.380..963A,2008MNRAS.390.1267T,2012MNRAS.426.3112H,2012ApJS..201...18M}. However galaxies are multi-component systems, 
and its simulation requires a robust model included special physical processes. So the main goal of this project is to
develop the numerical code for simulation of the evolution of disc galaxies including generation of spiral structure, physics 
of interstellar medium, formation of clouds and stars~\cite{2004ARA&A..42..211E,draine2010physics}.

This paper is organized as follows. In Section 2 the main numerical methods are described. Section 3 presents 
the results of tests. Section 4 describes the application of our code to the evolution of 
stellar-gaseous galactic disc. In Section 5 we summarize our results.

\section{Description of the code}

\subsection{Gas dynamics}

The gas dynamics equations for an ideal magnetized gas can be written in the conserved variables ${\bf U}$ in 
Cartesian coordinates
\begin{equation}
 {\bf U}  = \left[ 
  \rho ,\,\,
  \rho v_{x} ,\,\, 
  \rho v_{y} ,\,\, 
  \rho v_{z} ,\,\,
   E ,\,\,
   B_{x} ,\,\,
   B_{y} ,\,\,
   B_{z}  \right]^T. 
\label{eq:cons}
\end{equation} 
Then, the conservation laws can be written in a compact form 
\begin{equation}
\frac{\partial {\bf U}}{\partial t} + 
    \frac{\partial \bf F}{\partial x} + \frac{\partial \bf G}{\partial y}
 + \frac{\partial\bf H}{\partial z} = {\bf S},
\label{eq:cons_laws}
\end{equation}
where ${\bf F}$, ${\bf G}$, and ${\bf H}$ are vectors of fluxes in $x$, $y$, and $z$-directions, respectively and ${\bf S}$ is the source vector. 
The components of these vectors are 
\begin{equation}
 {\bf F} = \left[ \begin{array}{c}
  \rho v_{x} \\
  \rho v_{x}^{2} + P^{*} - B_{x}^{2} \\
  \rho v_{x}v_{y} - B_{x}B_{y} \\
  \rho v_{x}v_{z} - B_{x}B_{z} \\
  (E + P^{*})v_{x} - ({\bf B}\cdot{\bf v})B_{x} \\
  0 \\
  B_{y}v_{x} - B_{x}v_{y} \\
  B_{z}v_{x} - B_{x}v_{z} \end{array} \right]
{\bf G} = \left[ \begin{array}{c}
  \rho v_{y} \\
  \rho v_{y}v_{x} - B_{y}B_{x} \\
  \rho v_{y}^{2} + P^{*} - B_{y}^{2} \\
  \rho v_{y}v_{z} - B_{y}B_{z} \\
  (E + P^{*})v_{y} - ({\bf B}\cdot{\bf v})B_{y} \\
  B_{x}v_{y} - B_{y}v_{x} \\
  0 \\
  B_{z}v_{y} - B_{y}v_{z} \end{array} \right]
{\bf H} = \left[ \begin{array}{c}
  \rho v_{z} \\
  \rho v_{z}v_{x} - B_{z}B_{x} \\
  \rho v_{z}v_{y} - B_{z}B_{y} \\
  \rho v_{z}^{2} + P^{*} - B_{z}^{2} \\
  (E + P^{*})v_{z} - ({\bf B}\cdot{\bf v})B_{z} \\
  B_{x}v_{z} - B_{z}v_{x} \\
  B_{y}v_{z} - B_{z}v_{y} \\
  0   \end{array} \right]
\label{eq:z-flux}
\end{equation}
where $\Oo P^* = P + \frac{B^2}{2}$ and total gas energy $\Oo E = \frac{P}{\gamma - 1} + \frac{1}{2}\rho v^2 + \frac{B^2}{2} $. For pure gas dynamics we set ${\bf B}\equiv 0$. 

In multi-dimensional magneto-gas dynamics to obtain zero-value for the magnetic field divergence we apply the
Constrained Transport technique for magnetic field transport through computational domain~\cite{1966ITAP...14..302Y, 
1988ApJ...332..659E}. In this approach magnetic field strength is defined at faces of a cell, while other 
gas dynamical variables are defined at the center of a cell. 

We solve the equations (\ref{eq:cons_laws}) in Cartesian coordinates $(x,y,z)$. The center of each cell is 
located at the position ($x_i$, $y_j$, $z_k$). For each cell the faces being normal to the $x$ direction 
have coordinates $x_{i\pm 1/2}$ and sizes $\delta x$, $\delta y$ and $\delta z$. The gas dynamic variables (density,
momentum, energy) are volume-averaged and defined at the center of a cell. For example, the density value 
can be written as follow
\begin{equation}
\Oo \rho_{i,j,k} = \frac{1}{\delta x \delta y \delta z} \int^{x_{i+1/2}}_{x_{i-1/2}}
\int^{y_{j+1/2}}_{y_{j-1/2}}
\int^{z_{k+1/2}}_{z_{k-1/2}} \rho(x',y',z')dy'dz'\,.
\end{equation}
While the magnetic field components are surface-averaged and found at faces of a cell:
\begin{equation}
B_{x,{i-1/2,j,k}} = \frac{1}{\delta y \delta z} 
\int^{y_{j+1/2}}_{y_{j-1/2}}
\int^{z_{k+1/2}}_{z_{k-1/2}} B_x(x_{i-1/2},y',z')dy'dz'\,.
\end{equation}
To calculate the momentum and energy fluxes we apply the second-order cell-centered averages for magnetic field, 
e.g. for $B_x$:
\begin{equation}
\Oo B_{x,i,j,k} = \frac{1}{2} ( B_{x,i+1/2,j,k} + B_{x,i-1/2,j,k} ) \,.
\end{equation}

We use a finite-volume discretization for gas dynamics equations and finite-area discretization for magnetic field transport equations. In this case gas-dynamical variables at time $t+\delta t$ can be found as follow
\begin{equation}
{\bf U}_{i,j,k}^{n+1} = {\bf U}_{i,j,k}^{n} - \frac{\delta t}{\delta x} \left[{\bf F}_{i+1/2,j,k}^{n+1/2} - {\bf F}_{i-1/2,j,k}^{n+1/2}  \right]- \frac{\delta t}{\delta y} \left[{\bf G}_{i,j+1/2,k}^{n+1/2} - {\bf G}_{i,j-1/2,k}^{n+1/2}  \right]- \frac{\delta t}{\delta z} \left[{\bf H}_{i,j,k+1/2}^{n+1/2} - {\bf H}_{i,j,k-1/2}^{n+1/2}  \right]\,,
\end{equation}
where the vectors ${\bf F}_{i+1/2,j,k}^{n+1/2}$, ${\bf G}_{i,j+1/2,k}^{n+1/2}$, ${\bf H}_{i,j,k+1/2}^{n+1/2}$ are the time- and area-averaged fluxes through the $x$, $y$ and $z$-faces of a cell, correspondingly.

Using the Stokes' law for the magnetic field transport equation we obtain time-evolution of magnetic field equation:
\begin{equation}
B_{x,i-1/2,j,k}^{n+1} = B_{x,i-1/2,j,k}^{n} - \frac{\delta t}{\delta y} \left[ \E^{n+1/2}_{z,i-1/2,j+1/2,k} - \E^{n+1/2}_{z,i-1/2,j-1/2,k} \right] + \frac{\delta t}{\delta  z} \left[ \E^{n+1/2}_{z,i-1/2,j,k+1/2} - \E^{n+1/2}_{z,i-1/2,j,k-1/2} \right]\,,
\end{equation}
where $\E_x$, $\E_y$ and $\E_z$ are the components of electric field (or electro-magnetic force):
\begin{equation}
 \Oo \E = - {\bf v} \times {\bf B}\,.
\end{equation}
The approximation of the electro-magnetic force can be written as follow:
\begin{eqnarray}
 \Oo \E_{x,j-1/2,k-1/2} = \frac{1}{4} \left[ \E_{x,j-1/2,k} + \E_{x,j-1/2,k+1} + \E_{x,j,k-1/2} + \E_{x,j+1,k-1/2} \right] + \nonumber \\ 
 + \frac{\delta y}{8} \left[ \left( \frac{\partial \E_x}{\partial y} \right)_{j-1/4,k-1/2} - \left( \frac{\partial \E_x}{\partial y} \right)_{j-3/4,k-1/2} \right]
 + \frac{\delta z}{8} \left[ \left( \frac{\partial \E_x}{\partial z} \right)_{j-1/2,k-1/4} - \left( \frac{\partial \E_x}{\partial z} \right)_{j-1/2,k-3/4} \right]\,,
\end{eqnarray}
where the derivative of $\E_x$ for each grid cell face is computed by selecting the upwind direction according to 
the contact mode, namely,
\begin{equation}
\Oo \left(\frac{\partial \E_x}{\partial y}\right)_{k-1/2} = \left \{
\begin{array}{ll}
(\partial \E_x/\partial y)_{k-1} & \textrm{if } v_{z,k-1/2} > 0 \\
(\partial \E_x/\partial y)_{k} & \textrm{if } v_{z,k-1/2} < 0 \\
\frac{1}{2}\left[
\left( \frac{\partial \E_x}{\partial y} \right)_{k-1} + 
\left( \frac{\partial \E_x}{\partial y} \right)_{k}
\right] & \textrm{otherwise.}
\end{array}
\right .
\end{equation} 
Note that in 3D the expressions similar to the above are required to convert the $x$- and $y$-components
of the electric field to the appropriate cell corners. These expressions might be obtained directly cyclic permutation
of the $(x,y,z)$ and $(i, j, k)$.

Godunov-type methods are widely used for solving of hyperbolic equations. An idea is to use the Riemann solution 
for the decay of an arbitrary discontinuity. In this case it is assumed that the solution might be non-smooth and discontinuities might be in every computational cell. That is why, this kind of numerical schemes is a universal
tool for simulation of gasdynamical problems of evolution shocks and contact discontinuities. The main problem of Godynov-type methods is a necessity of solving system of nonlinear equations. Usually this leads to iterative procedure, which requires additional computational resources. To resolve this issue there are a lot of approximate solvers of Riemann problem. Our code includes two types of solvers for pure gas dynamics: Harten-Lax-van Leer (HLL) or Harten-Lax-van Leer-Contact (HLLC), and one for magneto-gas dynamics, namely, Harten-Lax-van Leer-Discontinuities (HLLD). 
These methods are described by~\cite{2005JCoPh.208..315M} in detail.  

An exact or approximate Riemann solver requires values for the conserved variables at left and right interfaces 
of each cell: $U_L = U_i-1/2$, $U_R = U_i-1/2$. These values can be reconstructed from cell centered values with 
second-order or third-order accuracy, on a choice. The interpolation of the values is carried out for primitive variables 
$w \equiv \{\rho, P, {\bf v}, {\bf B \}}$ as follow
\begin{equation}
dw_i = w_{i+1} - w_i\,,
\end{equation}
\begin{equation}
dW_i = K_2 \textrm{minmod}(dw_{i-1},dw_{i}) + K_1\textrm{minmod}(dw_{i},b dw_{i-1})\,,
\end{equation}
\begin{equation}
dW_{i+1} = K_2 \textrm{minmod}(dw_{i+1},dw_{i}) + K_1 \textrm{minmod}(dw_{i},b dw_{i+1})\,,
\end{equation}
and then $w_L = w_i + dW_i\,$ $w_R = w_{i+1} - dW_{i+1}\,,$ where $\Oo K_1 = \frac{1}{4}(1-\kappa)$, $\Oo K_2 = \frac{1}{4}(1+\kappa)$, $\Oo b = (3-\kappa)/(1-\kappa)$ and $\textrm{minmod}$ is a limiter function~\cite{1983JCoPh..49..357H}. The parameter $\kappa$ can be equal to $-1$,
$0$, or $1/3$, which represent second-order fully upwind, second-order upwind biased, and third-order upwind biased cases, respectively. Below we use $\kappa = 1/3$.

\subsection{$N$-body dynamics}

We use the "particle-in-cell" method for the dynamics of both dark matter halo ("live" halo) and stellar discs. 
To obtain the second-order accuracy on time we apply the "flip-flop" integrator. The main advantage of this 
approach is that it requires only one calculation of forces per time step for each particle. More details about
our $N$-body solver for galaxy dynamics can be found in~\cite{2001ARep...45..180K}.

\subsection{Poisson solver}
To take into account self-gravity of a gas we solve the Poisson equation with the gas dynamics conservation laws~(\ref{eq:cons_laws}):
\begin{equation}
\Oo \Delta \Psi(x,y,z) = 4\pi G \rho(x,y,z)\,.
\end{equation}
In our code the Poisson equation can be solved using three different methods: FFT-based, TreeCode~\cite{1986Natur.324..446B} 
and over-relaxation Gauss-Seidel method. To calculate self-gravity in $N$-body/gasdynamical problems we use regular Cartesian grid. For interpolation of particle density to a mesh we use the second order finite-volume interpolation, which conserves the total mass of the system with a good accuracy.

\section{Test problems}
Below we present several tests for our gasdynamical code in 1D, 2D and 3D.

\subsection{The Sod shock tube test}
\begin{figure}[t!]
\begin{minipage}[h]{0.63\linewidth}
\hskip-0.07\hsize\includegraphics[width=0.4\hsize]{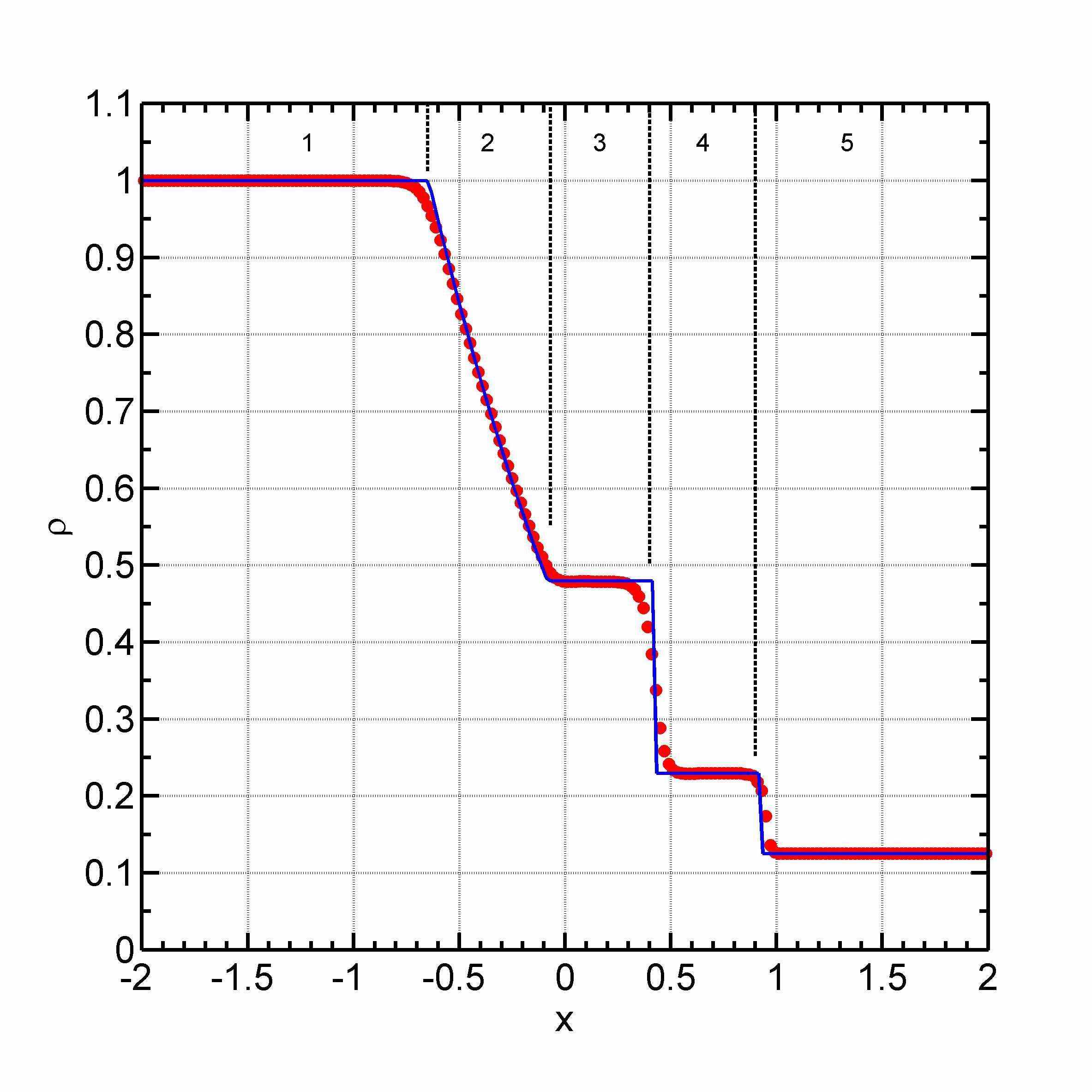}\includegraphics[width=0.4\hsize]{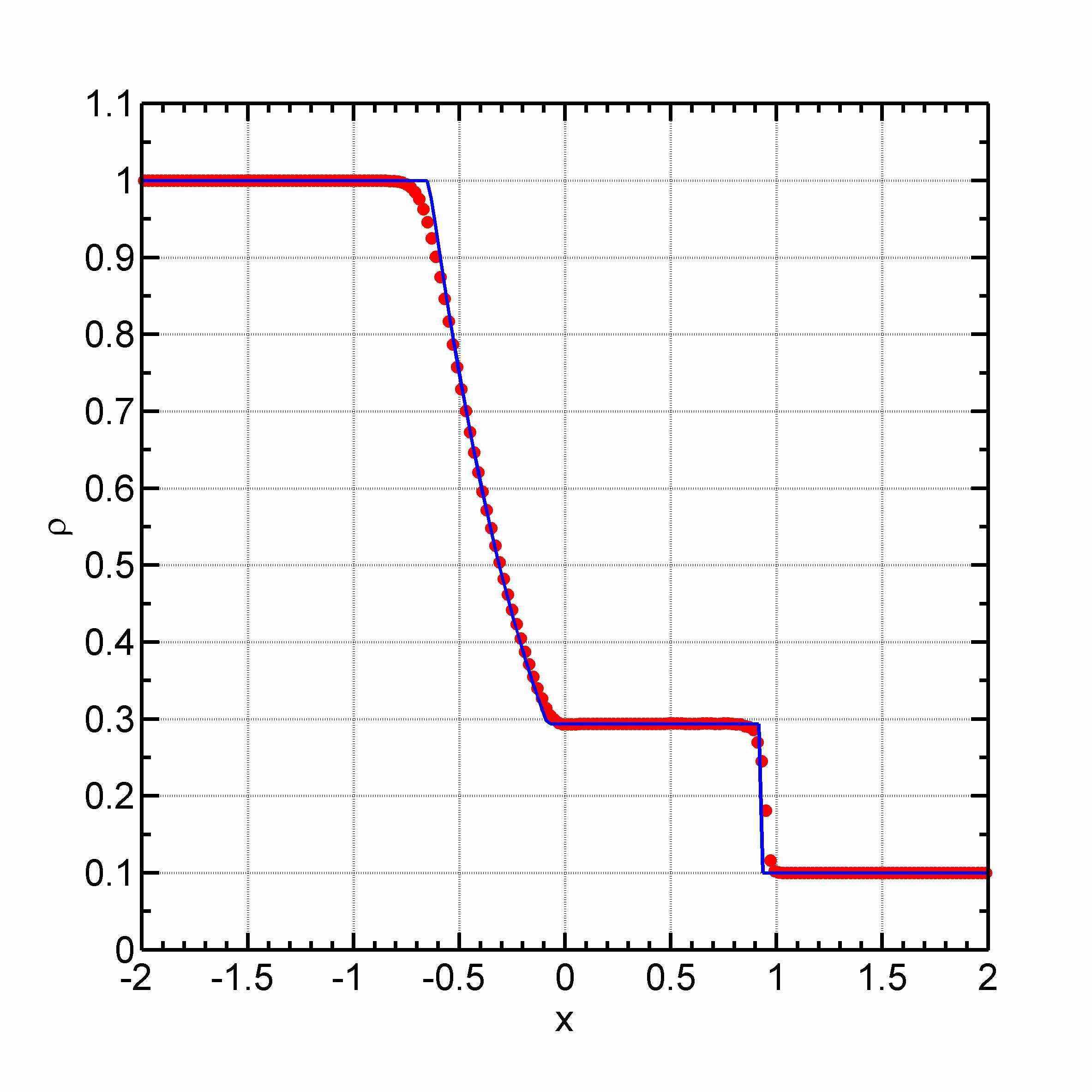}\includegraphics[width=0.4\hsize]{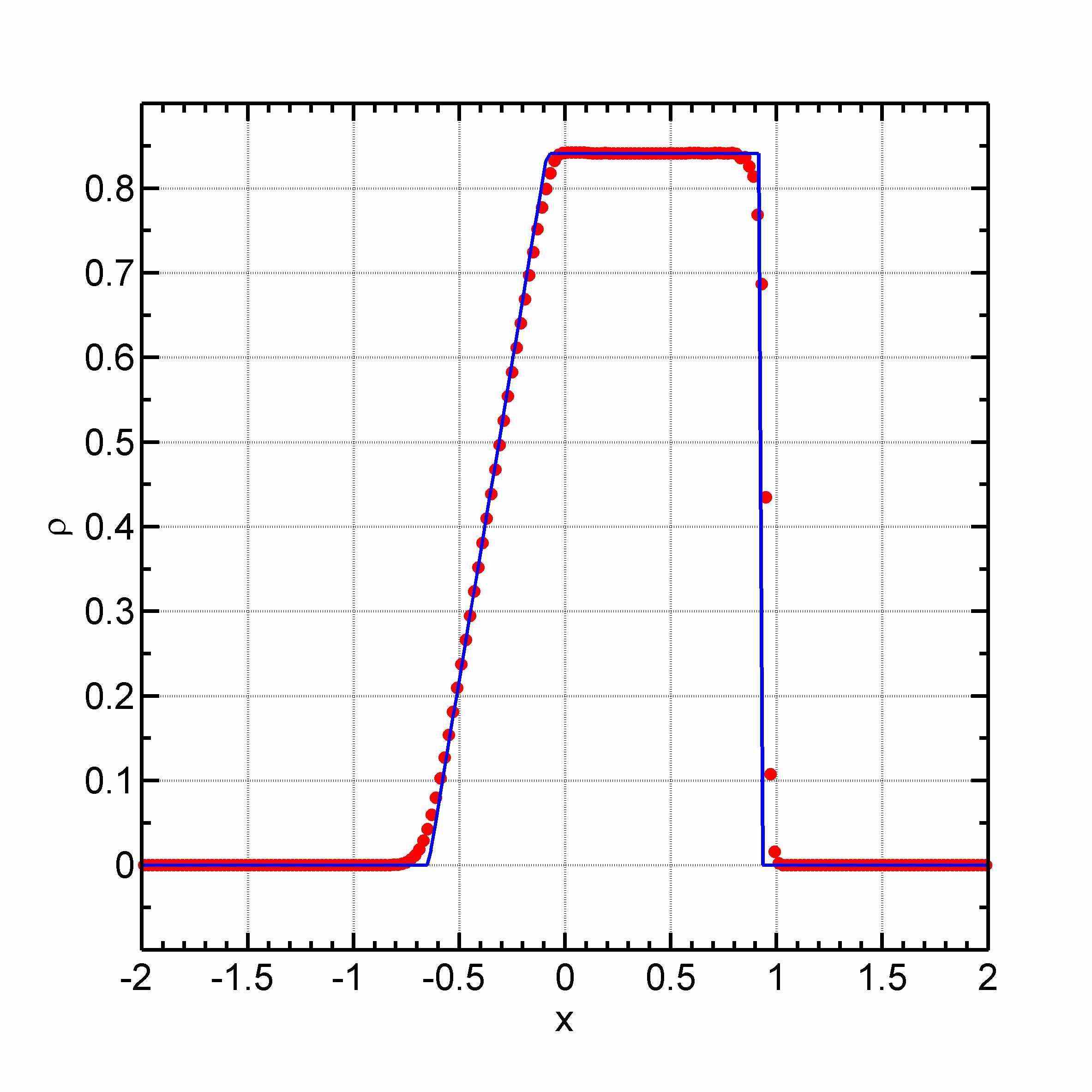}
\caption{The Sod shock tube. The exact solution is shown by black line, the other lines represent the distribution 
at $t=0.5$ with grid resolution $N=200$ (green), $N=500$ (red), $N=1500$ (blue).
One can find five regions in the solution: the regions 1 and 5 correspond to the initial unperturbed 
state of a gas, the rarefaction 
wave is found in the region 2, the regions 3 and 4 are separated by contact discontinuity, and the shock wave is located between
regions 4 and 5.
}
\label{fig::SOD}
\end{minipage}
\hfill 
\begin{minipage}[h]{0.3\linewidth}
\includegraphics[width=1\hsize]{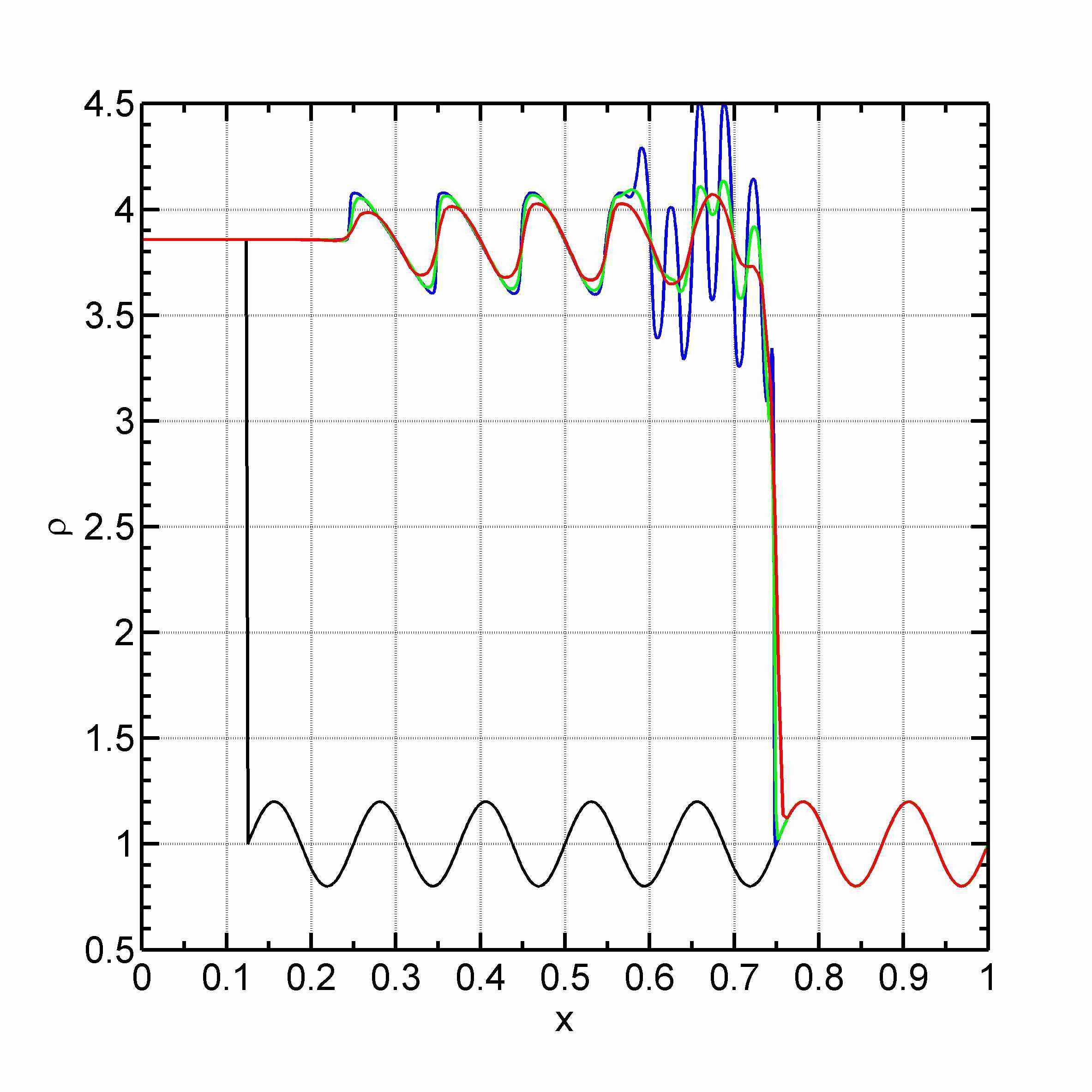}
\caption{The Shu-Osher test. The initial density distribution is shown by black line, the other lines represent the distribution 
at $t=0.18$ with grid resolution $N=200$ (green), $N=500$ (red), $N=1500$ (blue).}
\label{fig::ShuOsher}
\end{minipage}
\end{figure}

This is the most simple test demonstrated the formation of shock wave, contact discontinuity and rarefaction wave. 
The initial distribution of gasdynamical values is taken as usual: 
$\rho, v_x, P = \{1, 0, 1\}$ at $x<0$; $\rho, v_x, P = \{0.125,0, 0.1\}$ at $x>0$. Initially the velocity is equal 
to zero. 
Figure~\ref{fig::SOD} presents the result of numerical simulation for different resolution and the exact solution. 
One can see a good coincidence of the numerical solution with the exact one even for low resolution.

\subsection{The Shu-Osher problem}

The Shu-Osher problem tests a shock-capturing scheme's ability to resolve small-scale flow features. It shows the
numerical viscosity of the method. The initial distribution is: $\rho = 3.857143$, $p=10.33333$, $u=2.629369$ at $x<0.125$ and $\rho = 1 + 0.2\sin(16\pi x)$, $p=1$, $u=0$ otherwise, $\gamma$ equals 1.4.

Figure \ref{fig::ShuOsher} shows the initial density distribution (black line). We study the dependence on spatial
resolution: the case for $N=200$ is exactly smooth, whereas oscillations appear for $N=500$, for further increase of
resolution ($N=1500$) our result becomes closer to the exact solution~\cite{1989JCoPh..83...32S}.

\subsection{The bow shock simulation}

A bow shock usually forms due to supersonic motion of star (or planet) through interstellar medium~\cite{2013ApJ...764...19B} or a galaxy through the intercluster medium~\cite{2007MNRAS.380.1399R}, so that this is one of the most important astrophysical problem.  
To test our code we simulate a supersonic gas flow through potential well. The initial distribution of the parameters are homogeneous: $\rho = 1.0$, $P = 1.0$, $v_x = 1.0$, and $\gamma$ equals 1.4. The external potential well is set
in the form $\Oo \Psi(x) = - \Psi_0 \exp(-(x/x_0)^2)$, where $\Psi_0 = 5$ and $x_0 = 0.1$. The boundary conditions 
are free at the right and the steady inflow at the left. 

\begin{figure}[t!]
\begin{center}
\includegraphics[width=0.25\hsize]{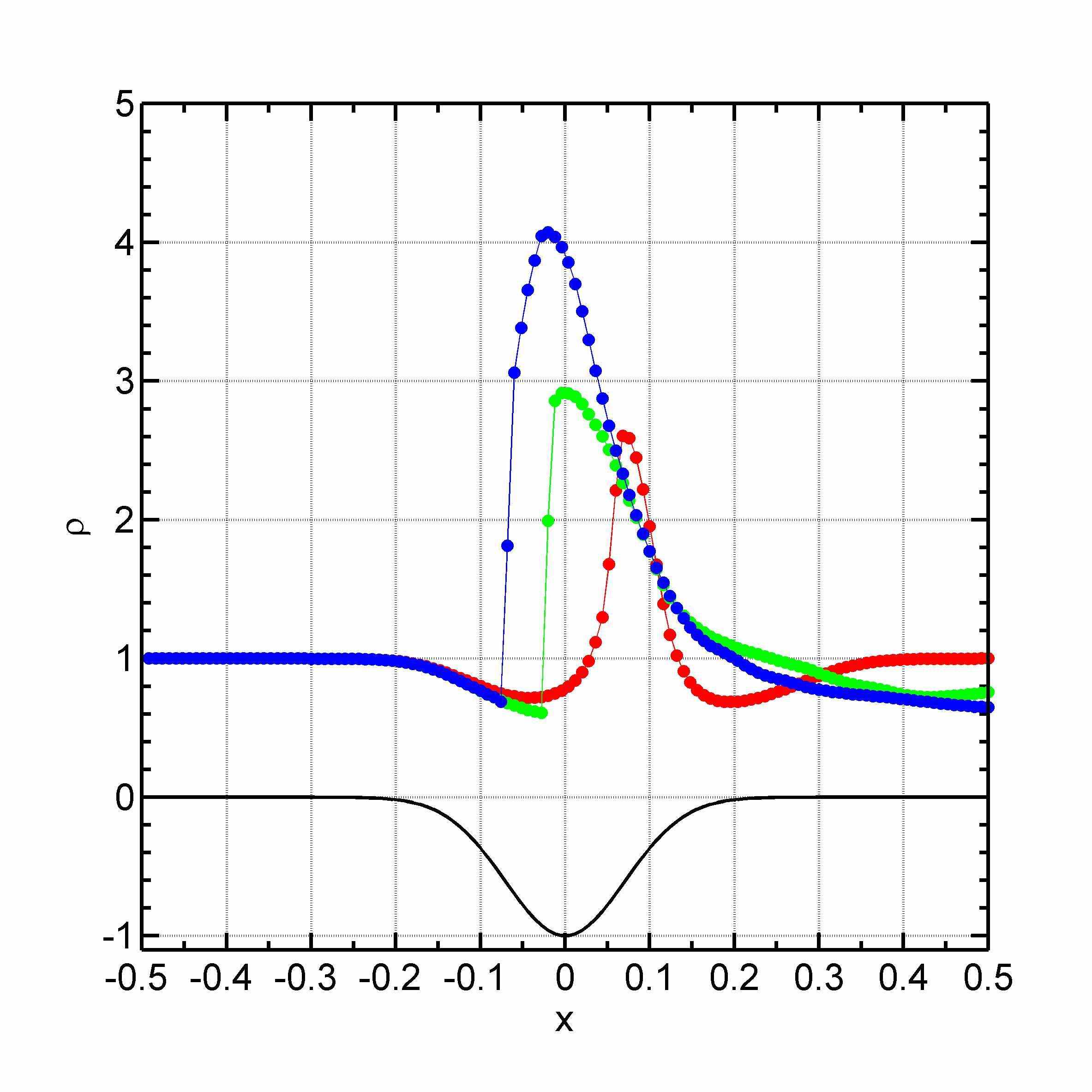}\includegraphics[width=0.25\hsize]{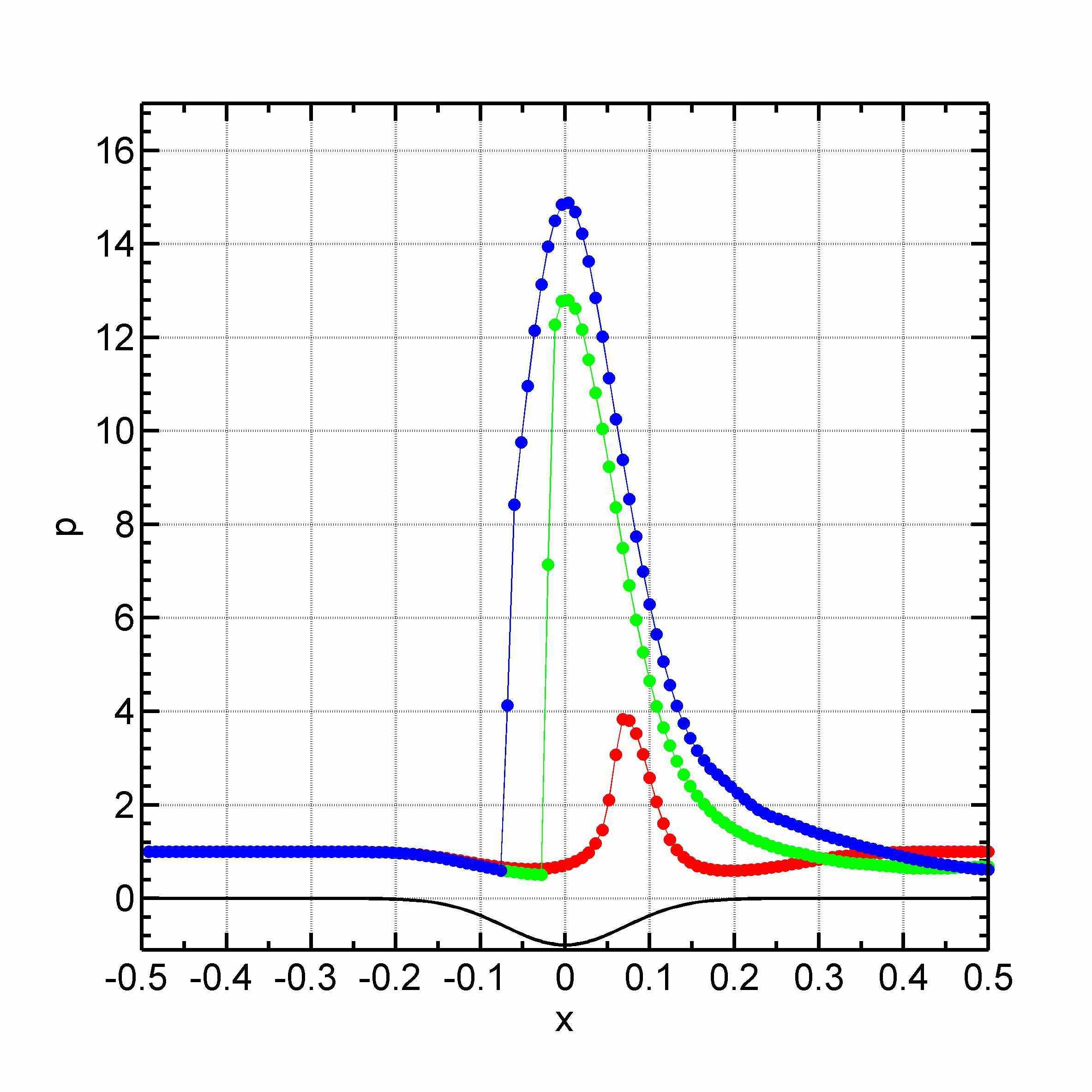}\includegraphics[width=0.25\hsize]{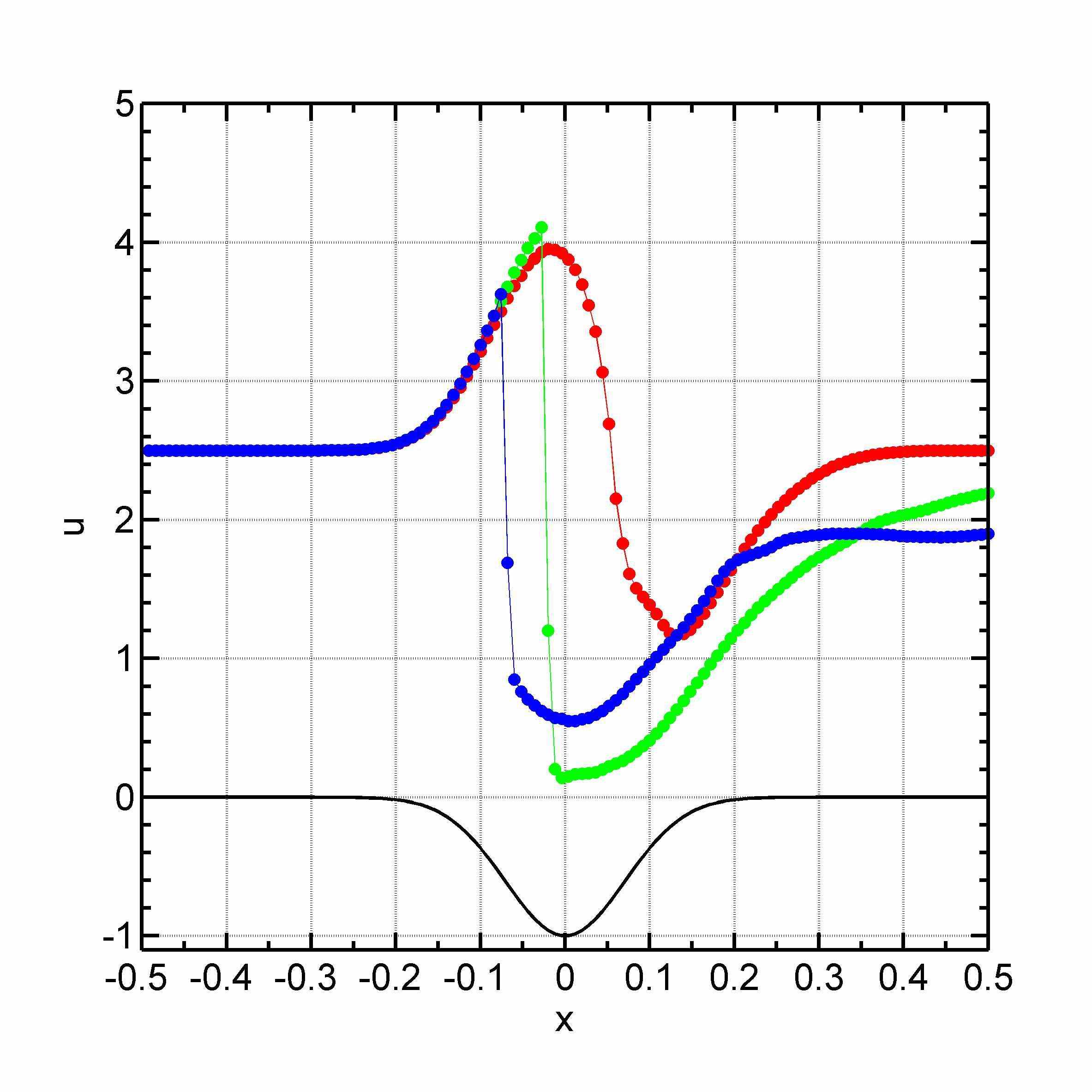}
\caption{The bow shock test. The density distribution is at $t = 0.051$~(red), $0.226$~(green) and $0.375$~(blue). 
The potential well $\Psi(x)/ \Psi_0$ is shown by black solid line.}
\label{fig::bowshock}
\end{center}
\end{figure}

At $t = 0.051$ the shock wave is formed on the far edge of the well (towards to the flow) due to supersonic falling 
of the gas to the potential well (red line). Note that this configuration is unstable. So the shock wave moves 
upstream through the potential well and after crossing the minimum of the potential it becomes steady-state at front edge of the potential well~(blue line).

\subsection{The Brio-Wu problem}

\begin{figure}[t!]
\includegraphics[width=0.9\hsize]{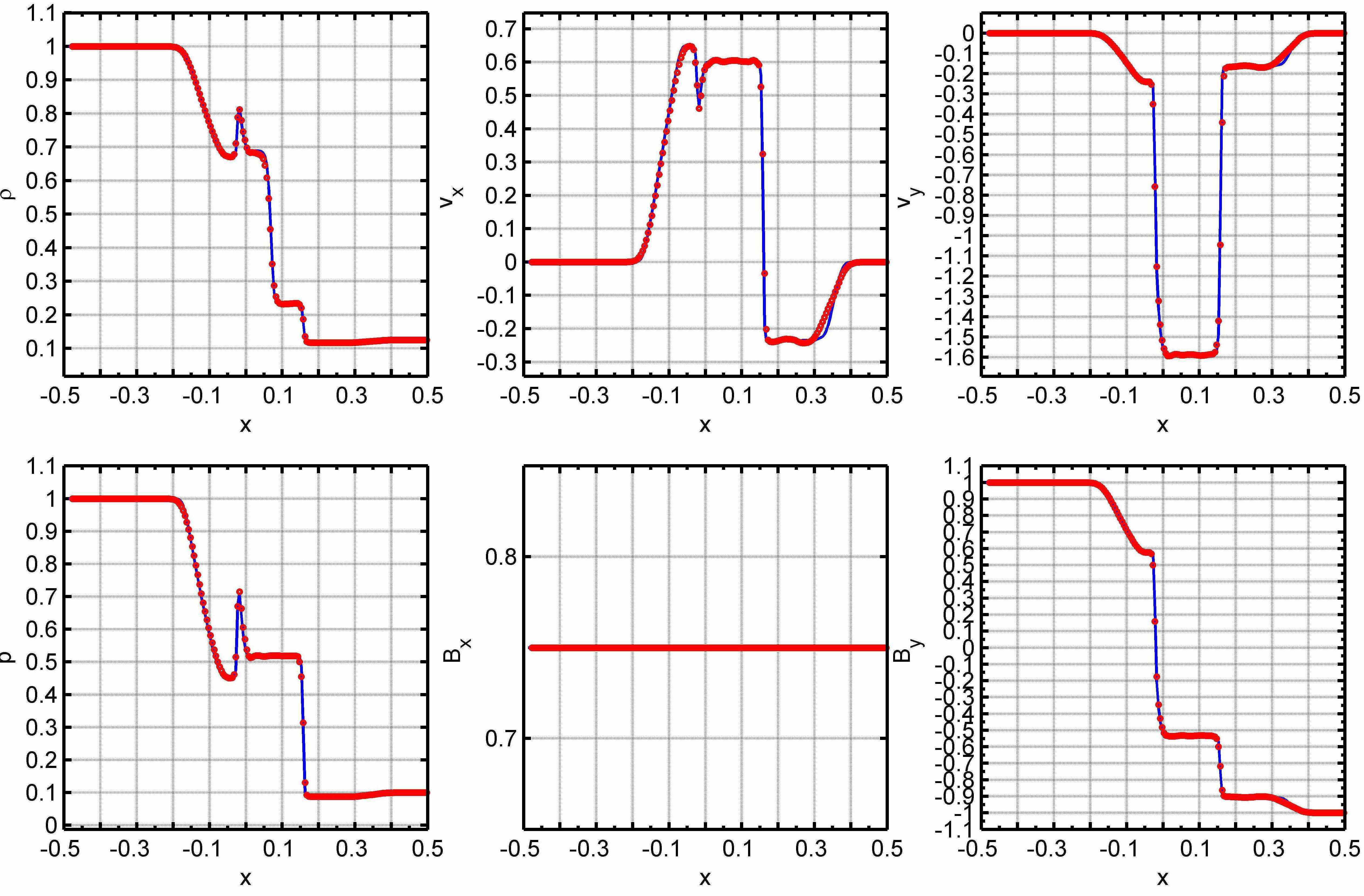}
\caption{The Brio-Wu problem. Our result is depicted by red dots, the result obtained in the Athena code is shown
by blue line.}
\label{fig::BW}
\end{figure}

The Brio-Wu test is one dimensional magneto-gas dynamics problem. The solution of this test consists of the fast rarefaction wave, that moves to the left, the intermediate shock wave, the slow rarefaction wave, the contact discontinuity, slow shock wave and another fast rarefaction wave, that moves to the right~(see figure~\ref{fig::BW}). 
The intermediate shock and slow rarefaction waves form a structure called the compound wave~\cite{1988JCoPh..75..400B}. The initial distribution for this test is: $\rho = 1$, $P=1$, $B_y=1$ at $x<0$, $\rho = 0.125$, $P=0.1$, $B_y=-1$ for $x>0$. The longitudinal component of magnetic field $B_x = 0.75$ is constant over the grid.

This test (and the next one, the Ryu-Jones problem) is compared with the results obtained by the Athena code~\cite{2008ApJS..178..137S}.
For both simulations we set the same spatial resolution $N=200$~(figure~\ref{fig::BW}). One can see a good agreement between 
the results, but the artificial oscillations can be found for the longitudinal velocity component for both solutions.

\subsection{The Ryu-Jones problem}

The Ryu-Jones problem tests the rotation of the magnetic field components. The solution consists of two fast shock waves with velocities 1.22 and 1.28 Mach numbers, two slow shock waves with 1.09 and 1.07 Mach numbers, two rotational and one contact discontinuities~\cite{1995ApJ...442..228R}. 

\begin{figure}[t!]
\includegraphics[width=0.9\hsize]{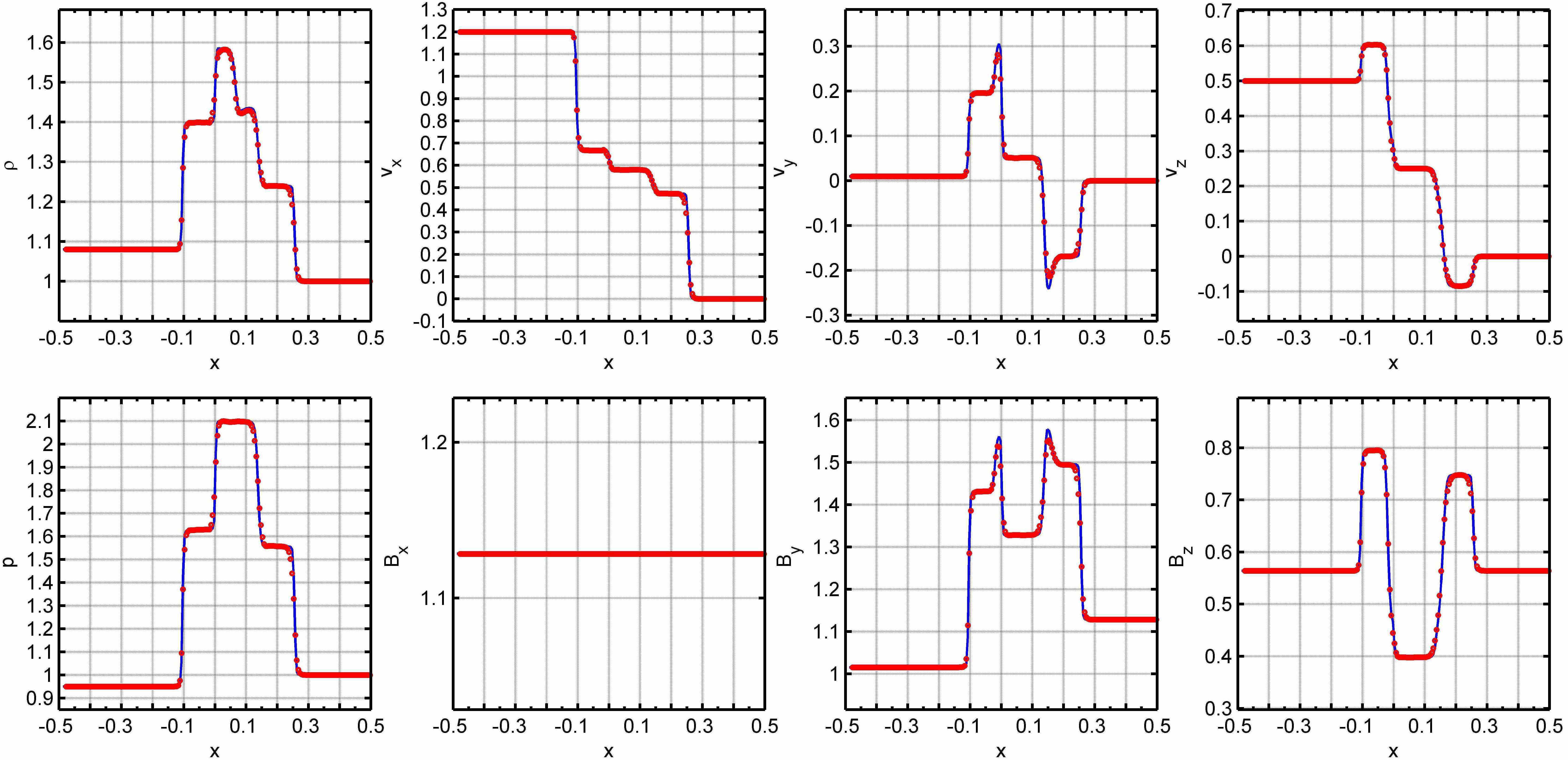}
\caption{The Ryu-Jones problem. Our result is depicted by red dots, the result obtained in the Athena code is shown
by blue line.}
\label{fig::RJ}
\end{figure}

The initial distribution is: $\rho = 1.08$, $P=0.95$, $v_x = 1.2$, $v_y=0.01$, $v_z=0.5$, $B_x=4/\sqrt{4\pi}$, $B_y=3.6/\sqrt{4\pi}$, $B_z=2/\sqrt{4\pi}$ at $x<0$, and $\rho = 1$, $P=1$, $v_x = v_y= v_z=0$, $B_x=4/\sqrt{4\pi}$, $B_y=4/\sqrt{4\pi}$, $B_z=2/\sqrt{4\pi}$ at $x>0$. One can see a good agreement between results obtained both
in our code and in the Athena code. However the peaks of $v_y$ and $B_y$ in our simulation are smoother than these obtained in the Athena code~\cite{2008ApJS..178..137S}.  

\subsection{The rotor problem}

\begin{figure}[t!]
\begin{center}
\includegraphics[width=0.9\hsize]{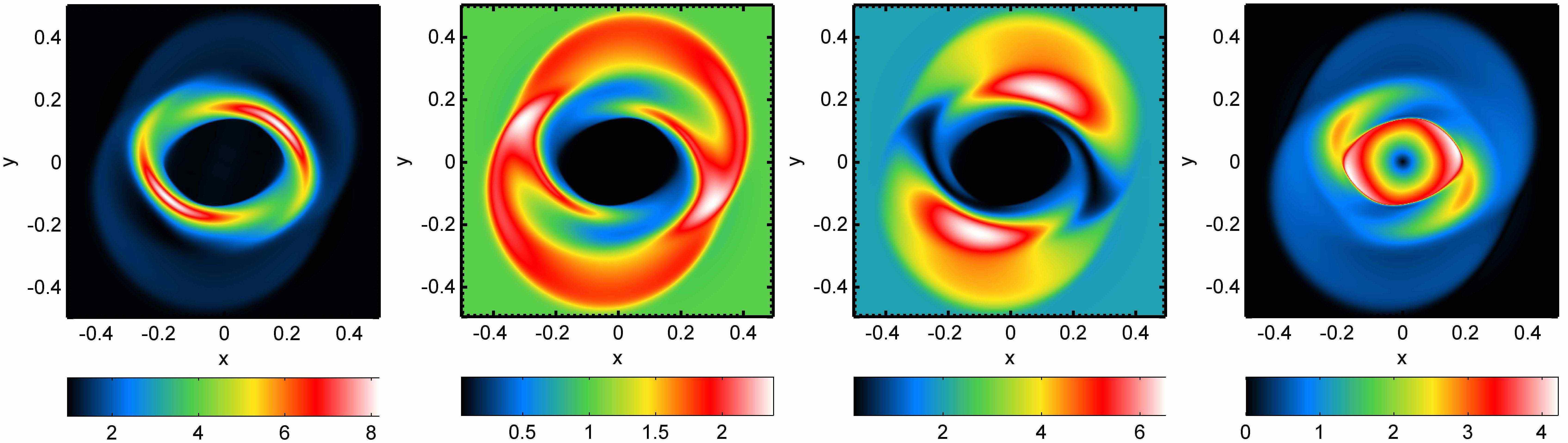}
\end{center}
\caption{The rotor problem. The maps of density, pressure, magnetic pressure and Mach number are shown from left to
right correspondingly.}
\label{fig::rotor}
\end{figure}

This test demonstrates fast rotation of the cylinder in the non-moving medium with homogeneous magnetic field. 
Initially there is a disc in the center of computational domain: $\rho = 10$, $v_x=-v_0y/r_0$ and $v_y=v_0x/r_0$ at
$r<r_0=0.1$; $\rho = 1+9f(r)$, $v_x=-v_0 f(r) y/r_0$, $v_y=v_0 f(r) x/r_0$ at $r_0<r<r_1=0.2$, and $\rho = 1$ at $r>r_1$;  $P=1$ and $B_x=5/\sqrt{4\pi}$ are constant in the whole computational domain. 
Here the function $f(r) = (r_1-r)/(r_1-r_0)$ is the linear interpolation of the gas density and velocity.
This configuration is strongly unstable, because the centrifugal forces are not balanced. A rotating gas is redistributed gradually within the computational domain capturing the stationary gas. Under these conditions
the magnetic field keeps the rotating material in the flattened form (Figure~\ref{fig::rotor}).

\subsection{The Sedov-Taylor blast-wave}

\begin{figure}[t!]
\includegraphics[width=0.9\hsize]{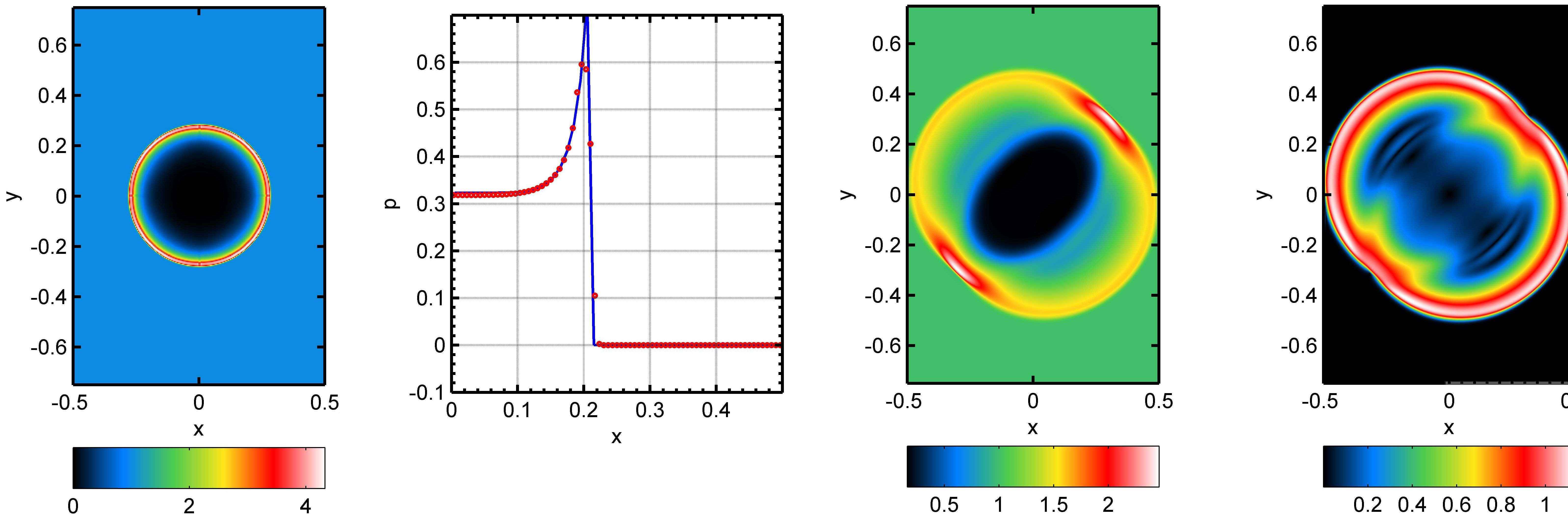}
\caption{The Sedov-Taylor blast-wave. The 2D density map for pure gas dynamics (left panel) and the radial pressure
profile at $t=0.1$ for our simulations (red dots in left middle panel) compared with the exact solution (blue line). 
The 2D density and magnetic pressure maps at $t=0.1$ for magneto-gas dynamics (two right panels).}
\label{fig::sedov}
\end{figure}

One of the most well-known self-similar solution describes a strong shock wave originated from the point explosion. 
For zeroed magnetic field a numerical solution can be compared with the exact solution. In 2D we simulate a strong
explosion with the following parameters: $\rho=1$, $P=10^{-5}$ are set in the whole computational domain,
$P=\pi^2r_0^2/(\gamma-1)$ at $r<r_0$. We set $r=0.01$ and take a computational domain $[-0.5,0.5]\times[-0.75,0.75]$
and number of cells $512\times 768$. 
Figure~\ref{fig::sedov} shows the 2D density map (left panel) and the radial pressure profile for our simulations
(red dots in left middle panel) compared with the exact solution (blue line). One can see a good agreement between
numerical data and analytic curve. 

We consider a blast wave in a magnetized medium. At $t=0$ we set $B_x = B_y = 1/\sqrt{8\pi}$. 
Despite of the symmetric initial distribution the shock front expands larger along the lines of the magnetic 
field.

\subsection{The Kelvin-Helmholtz instability}

Shear instabilities are very common in astrophysical objects, so that numerical algorithms should 
reproduce such inabilities well. Here we simulate the evolution of two gas layers moving with different
velocities and separated initially by a contact discontinuity: $\rho = 1$, $v_x = -0.5$ at $|y|<0.25$ 
and $\rho = 2$, $v_x = 0.5$ otherwise, and $P=2.5$ everywhere. We add a random perturbation of the velocity 
with the amplitude $0.001$. The grid size is $512\times 512$.

\begin{figure}[t!]
\includegraphics[width=1.0\hsize]{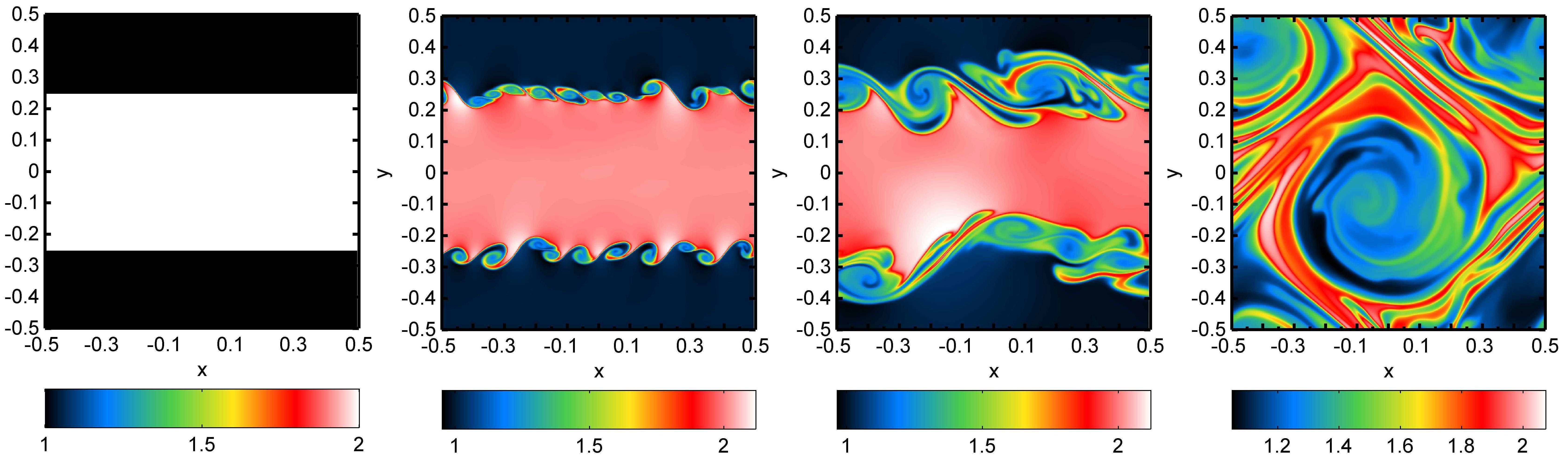}
\vskip -0.0\vsize
\includegraphics[width=1.0\hsize]{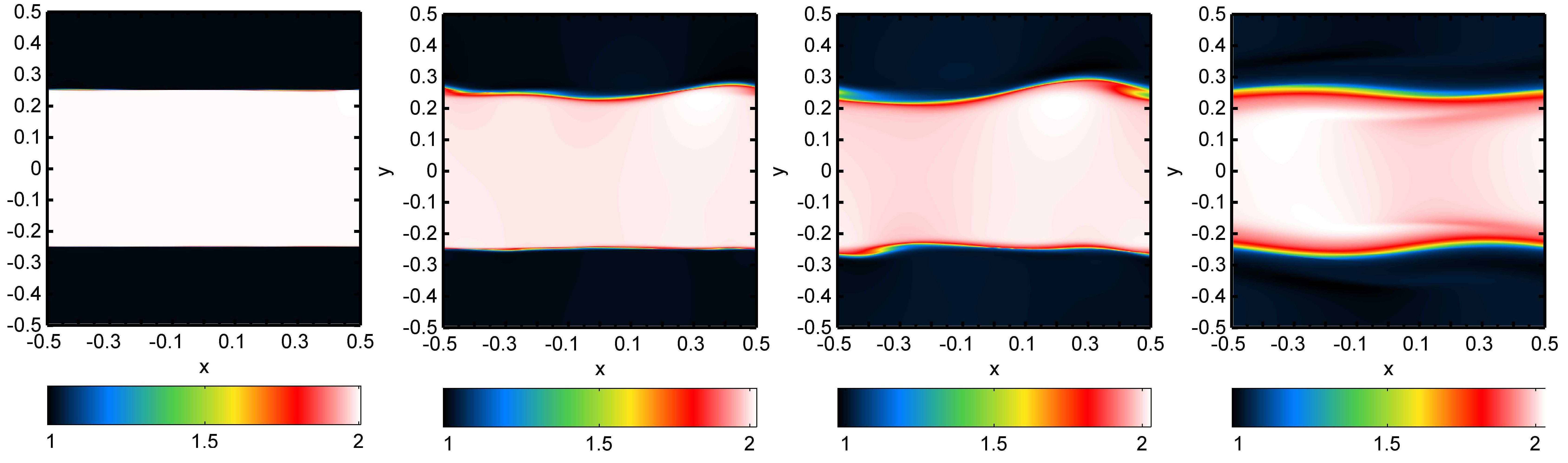}
\caption{The Kelvin-Helmholtz instability. The gas density distributions at $t=0, 0.1, 4, 12$ (from left to
right) in the pure gas dynamics (top row of panels) and the same, but in the presence of magnetic field (bottom 
row of panels).}
\label{fig::KH}
\end{figure}

Figure~\ref{fig::KH} shows the evolution of gas density for pure gas dynamics (upper row of panels) and 
magneto-gas dynamics (bottom row of panels). One can note the turbulent flow formation due to shear instability 
at early times, which produces large scale vortexes at further nonlinear stage. The presence of magnetic field 
strongly changes the picture:  a small scale perturbations disappear for initial longitudinal magnetic field $B_x = 0.5$~(see bottom line in Figure~\ref{fig::KH}), whereas a large scale oscillation of a gas along magnetic field
can be seen.  

\subsection{Spherical collapse of gravitating gas}

To test a gravity solver we use a standard 3D cosmological problem -- the collapse of a gaseous sphere with 
initial density profile $\rho(r) = 1/r$. We set the cubic computational domain $(x,y,z)\in [-1,1]$ with cell
number $100^3$. Figure~\ref{fig::collapse} shows the evolution of the density profile. One can see that dense 
core forms due to homogeneous collapse and strong shock wave moves outwards (see right panel of the Figure). 

\subsection{Thermodynamical module}

To study the evolution of the interstellar medium in galaxies we develop a module of thermal processes. In this module we can switch between tabulated cooling/heating rates and calculation of rates for main chemical species  in the interstellar medium. For the former we can use tables of cooling rates in the temperature range $10$~K$<T<10^8$~K calculated 
by \cite{1993ApJS...88..253S,2011MNRAS.414.3145V, 2013MNRAS.431..638V}. In the latter we  calculate the cooling and heating rates produced by specific emission processes and in the temperature range  $T< 2\times 10^4$~K, that is of a great importance for studying thermal state of the interstellar medium (see~\cite{1978ppim.book.....S}  for instance). We can investigate the thermal evolution of the interstellar medium for different abundances of heavy elements (metallicities). Our thermal block includes following cooling processes: cooling due to recombination and collisional
excitation and freeâ-free emission of hydrogen~\cite{1992ApJS...78..341C}, molecular hydrogen cooling~\cite{1998A&A...335..403G}, cooling in the fine structure and metastable transitions of carbon, oxygen and silicon~\cite{1989ApJ...342..306H}, energy transfer
in collisions with the dust particles~\cite{2003ApJ...587..278W} and recombination cooling on the dust\cite{1994ApJ...427..822B}. The heating rate takes into account photoelectric heating on the dust particles~\cite{1994ApJ...427..822B,2003ApJ...587..278W}, heating due to H$_2$ formation on dust and the H$_2$ photodissociation~\cite{1979ApJS...41..555H} and  the ionization heating by cosmic rays~\cite{1978ApJ...222..881G}. For multi-level atoms the cooling rates are obtained
from the level population equation assuming the optically thin steady-state regime~(see e.g.~\cite{1989ApJ...342..306H}).

\begin{figure}[t!]
\begin{minipage}[h]{0.49\linewidth}
\begin{center}
\includegraphics[width=1\hsize]{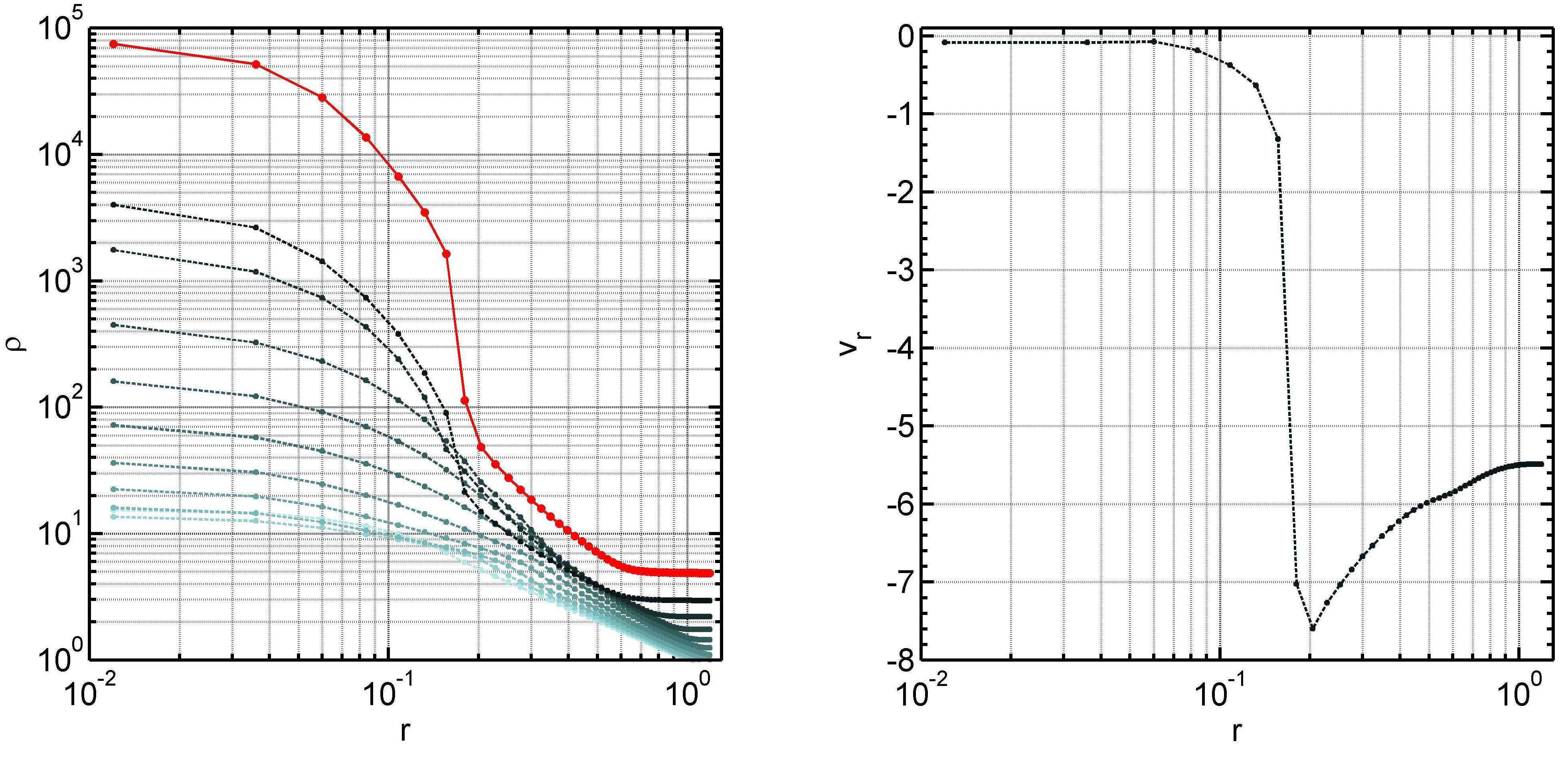}
\end{center}
\caption{The density profile for $t=0$ and $t=1$ is depicted by cyan and black lines, respectively, 
the pressure profile for $t=1$ is shown by red line (left panel). The velocity profile for $t=1$ is 
shown at right panel.}\label{fig::collapse}
\end{minipage}
\hfill
\begin{minipage}[h]{0.49\linewidth}
\includegraphics[width=1\hsize]{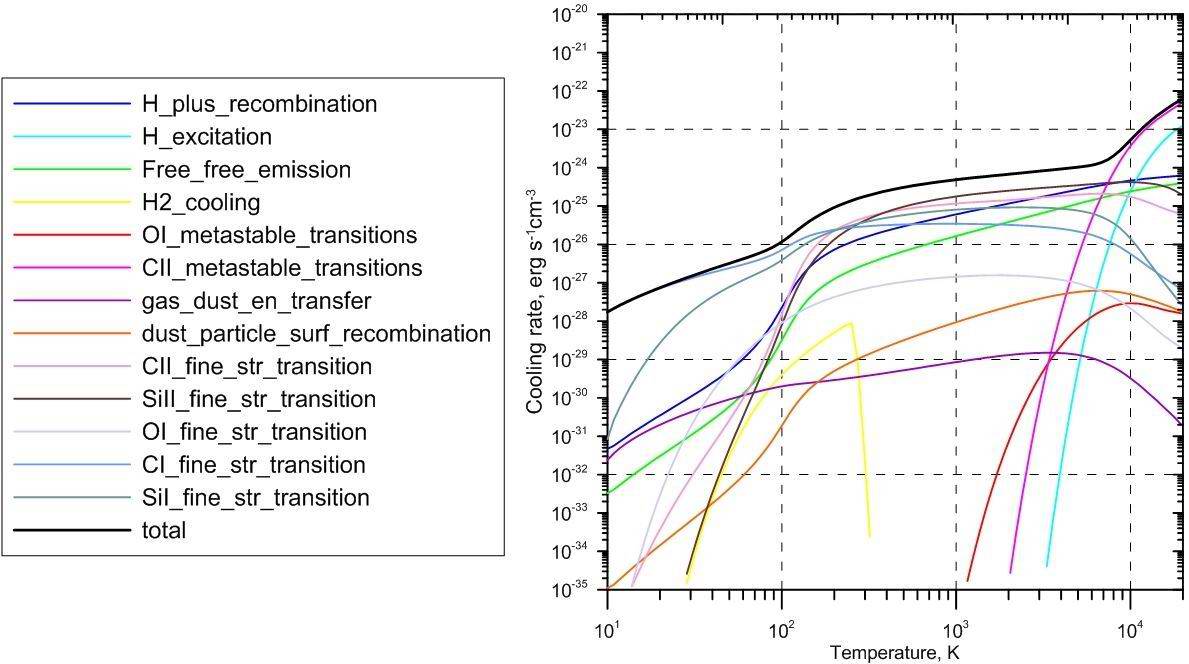}
\caption{Cooling functions, $\Lambda / n^2_H$, for solar metallicity. The total cooling rate is depicted by black line.}\label{fig::cooling}
\end{minipage}
\end{figure}
Figure~\ref{fig::cooling} presents cooling rates in the low temperature ($T<2\times 10^4$~K) range adopted for our model of the interstellar medium.

\subsection{$H_2$ kinetics}

Molecular hydrogen (H$_2$) regulates star formation in galaxies~\cite{1978ppim.book.....S}, then to study the evolution of the interstellar medium we need to take into consideration the H$_2$ kinetics. Because of significant difference between
dynamic and chemical timescales the H$_2$ kinetics is nonequilibrium and to get a source term in equation~(\ref{eq:cons_laws}) we solve a system of ordinary differential equations for the following chemical species: H, H$^+$, H$_2$. Molecular hydrogen is formed on the surface of dust grains and dissociated by ultraviolet Lyman-Werner photons and cosmic rays. Because of strong absorption and scattering of ultraviolet photons in the neutral hydrogen and dust to
mimic radiative transfer and H$_2$ self shielding we use a simplified approach introduced by~\cite{2008MNRAS.391..844D} for calculation of neutral and molecular hydrogen column densities. 

\subsection{Star formation implementation}
Stellar feedback effects can significantly influence on the galactic evolution. However, there is no consensus about the implementation of the star formation into gas dynamics, that is clearly demonstrated in ~\cite{2012MNRAS.423.1726S}. Their results of the galaxy evolution simulations within the $\Lambda$CDM paradigm strongly depend on the star formation and feedback recipes: stellar mass, size, morphology and gas content of the galaxy at $z=0$ vary significantly due to the different implementations of star formation and feedback. Despite this problem  we include the star formation effects using an intuitive approach, that partly based on the well-known method offered in~\cite{1997MNRAS.284..235Y}. 

To form a star or in general stellar particle we need to find cells in a computational domain, which satisfy to 
our conditions for star formation. These conditions can be more or less sophisticated and sometimes may have intricate nature. Usually the following criteria for the stellar particle creation in a given cell are considered: 
the surface density should be greater than the threshold for star formation $\Sigma_t$ and simultaneously the 
temperature should be less than some critical level $T_t$. In the simulations presented below we have assumed
$\Sigma_t>10^4$~cm$^{-3}$ and $T_t\le 10^2$~K. If these conditions are realised in a given cell, then we create a test 
stellar particle. The Salpeter initial mass function is assumed for each particle, which usually consists of 
$10^3-10^4$ stars. The initial velocity of this test particle is taken to be the same as its host cell. The 
kinematics of test particles are followed by the second-order integrator mentioned in Section~2.2. We take 
into account several feedback effects: stellar winds from massive stars, radiative pressure, supernova explosions, 
and stellar mass loss by low-mass stars~\cite{2013ApJ...770...25A}. Metals ejected by supernovae and lost by low-mass
stellar population can strongly change chemical composition and thermal evolution of a gas. Because of local character
of enrichment process the re-distribution (mixing) of metals is of great importance for further star formation
in galaxies~\cite{2009AstBu..64..317V, 2012ARep...56..895V,2004ARep...48....9D, 2002ApJ...581.1047D}.


\section{Simulation of the multicomponent galactic disc}

\begin{figure}[t!]
\begin{center}
\includegraphics[width=0.25\hsize]{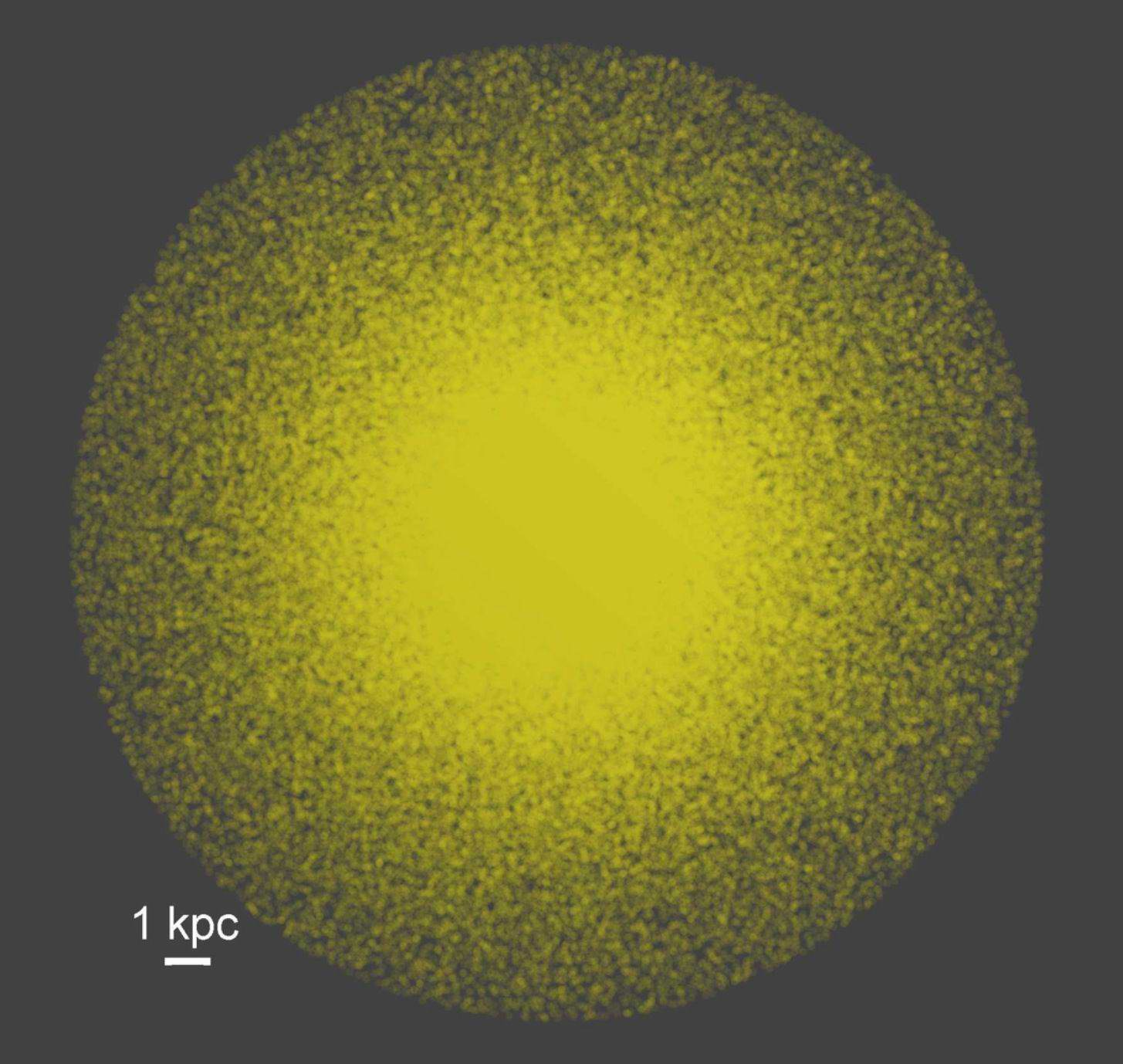}\includegraphics[width=0.25\hsize]{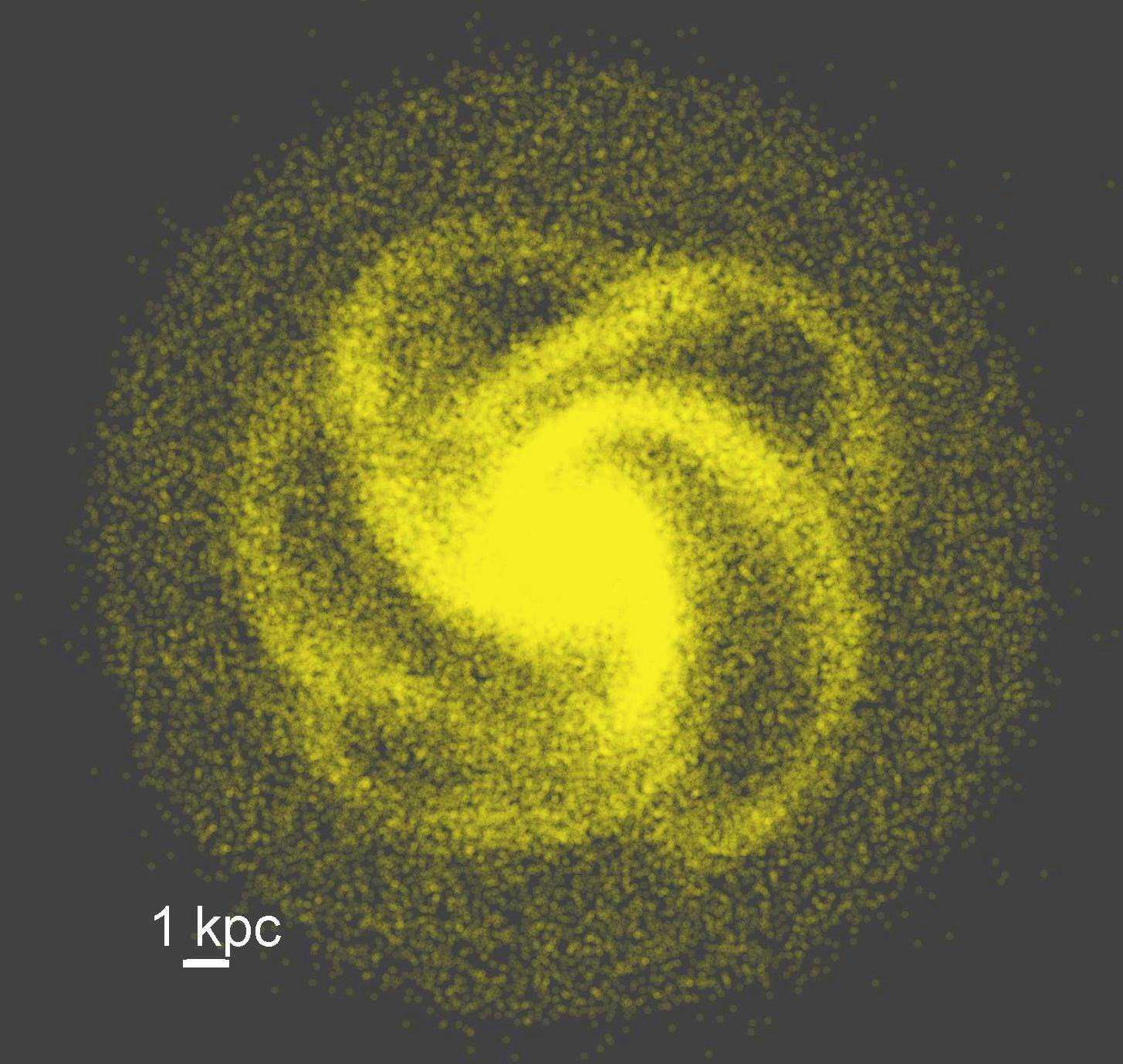}\includegraphics[width=0.25\hsize]{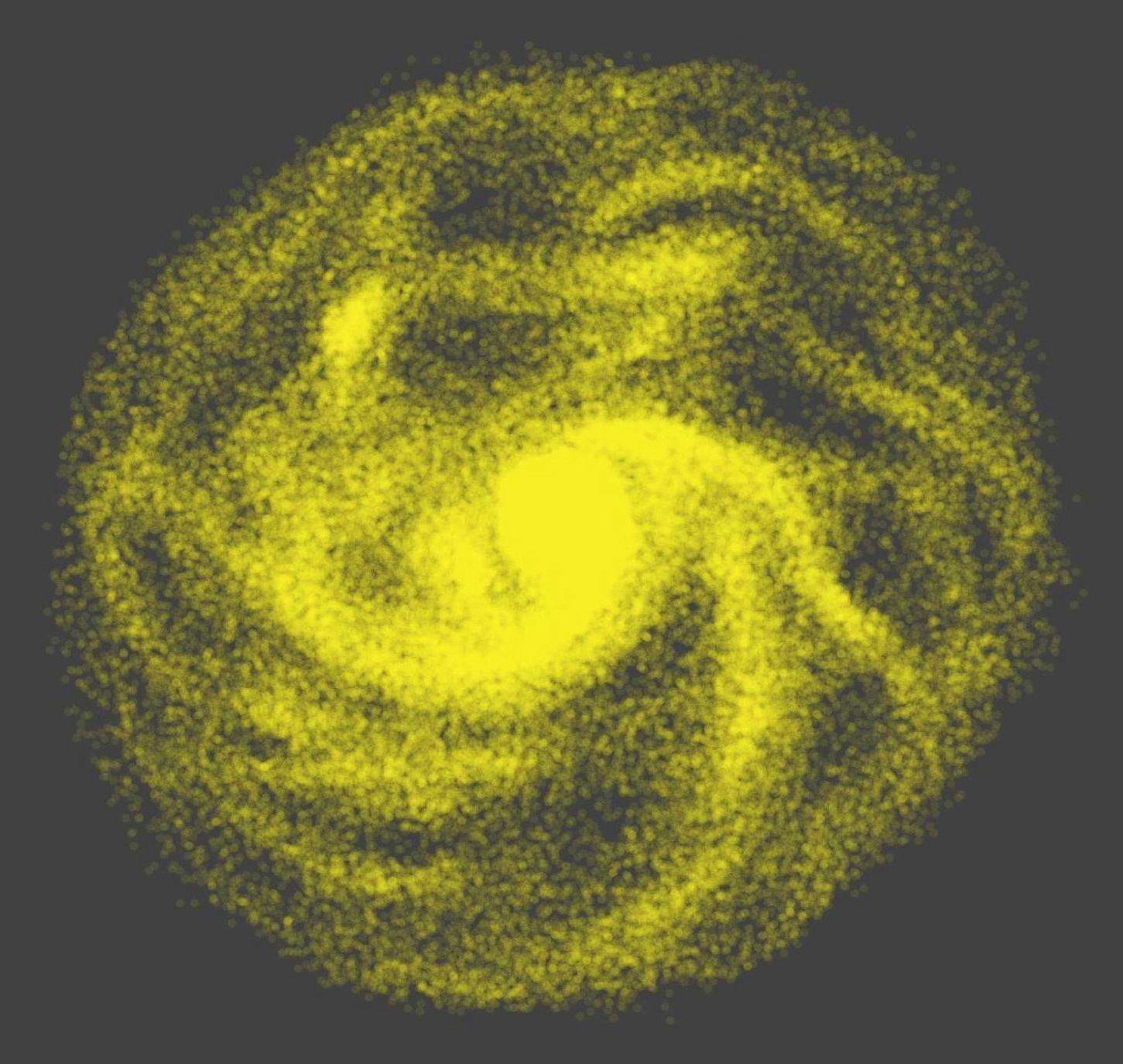}\includegraphics[width=0.25\hsize]{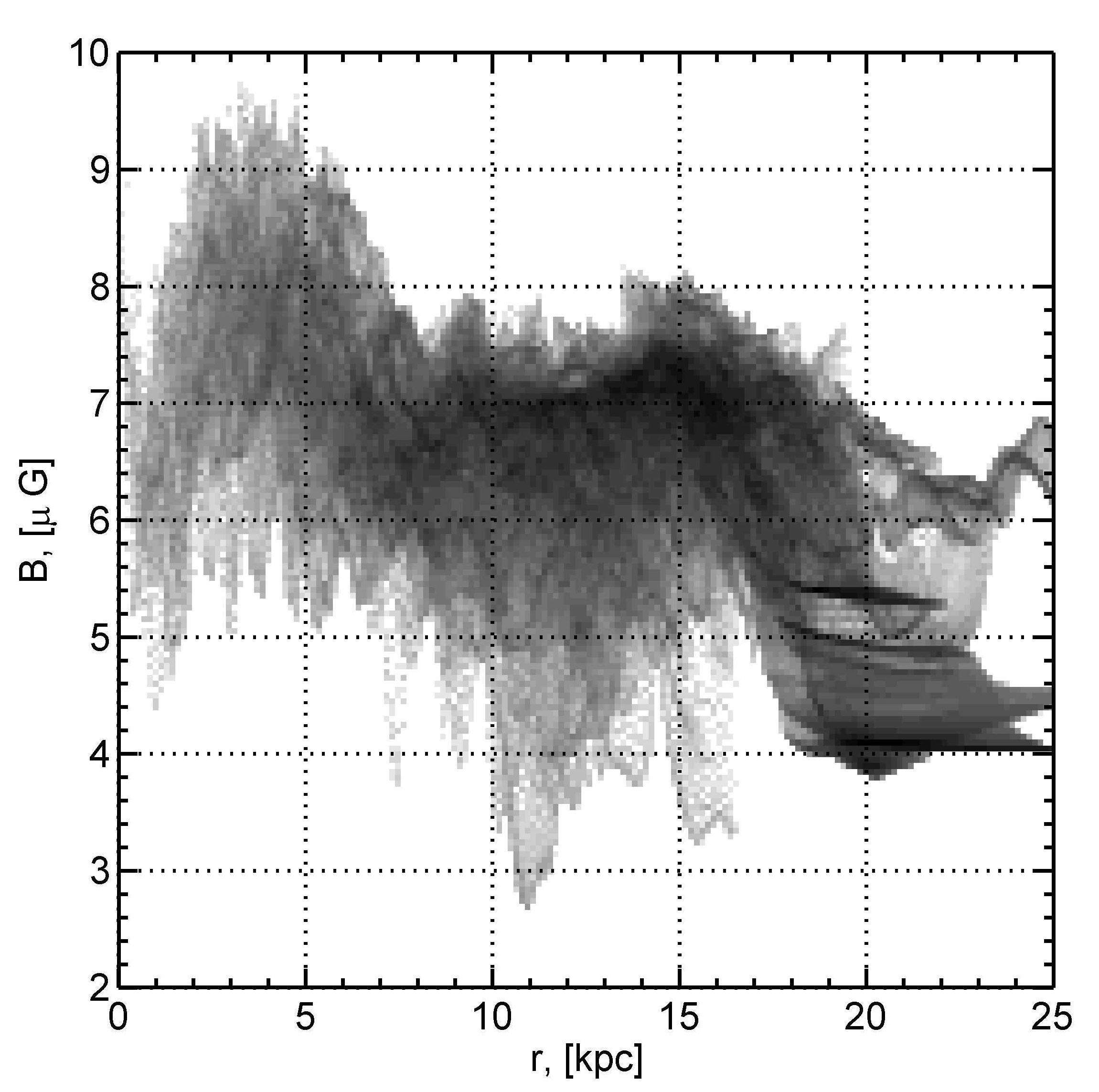}\\
\includegraphics[width=0.25\hsize]{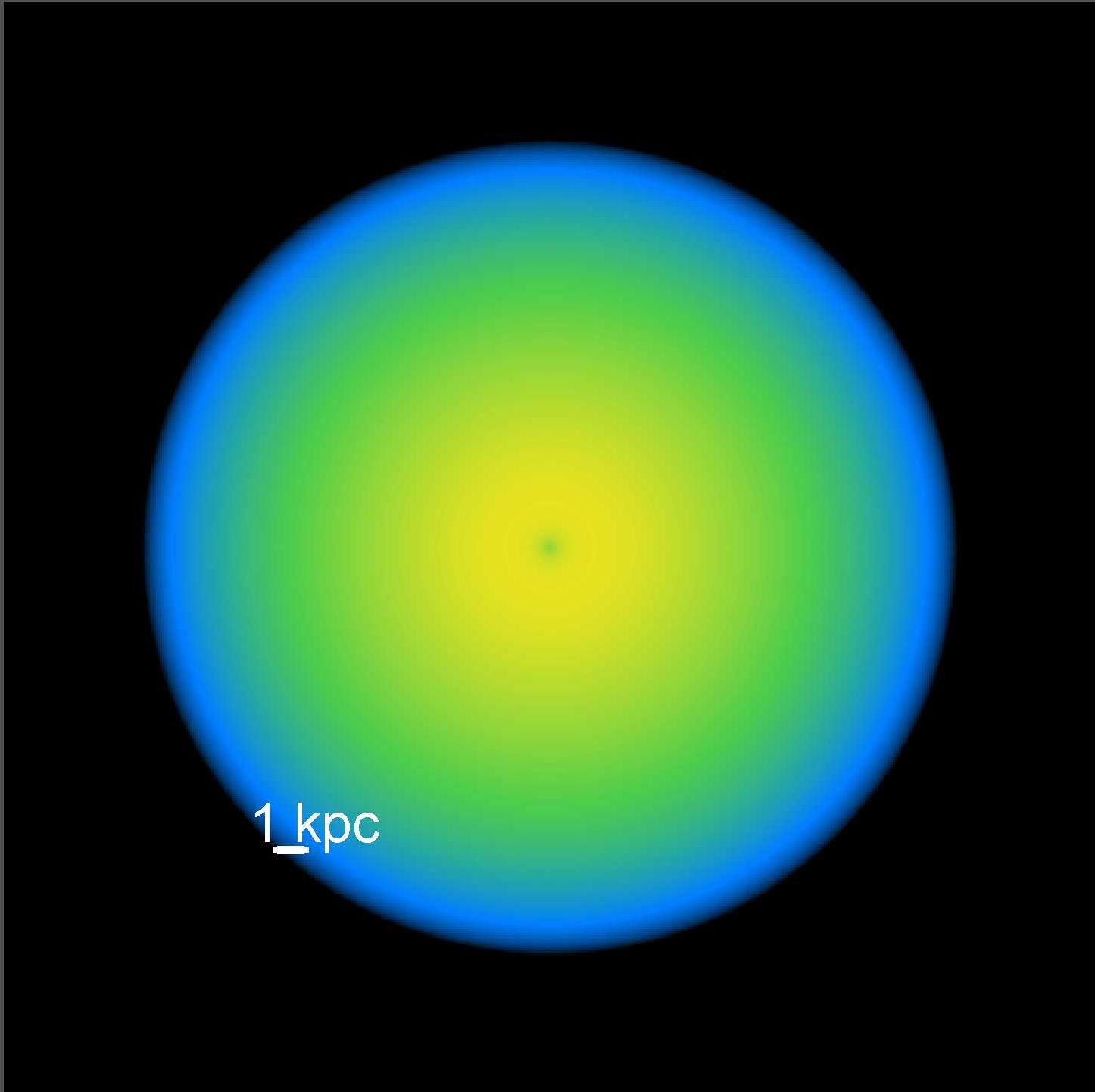}\includegraphics[width=0.25\hsize]{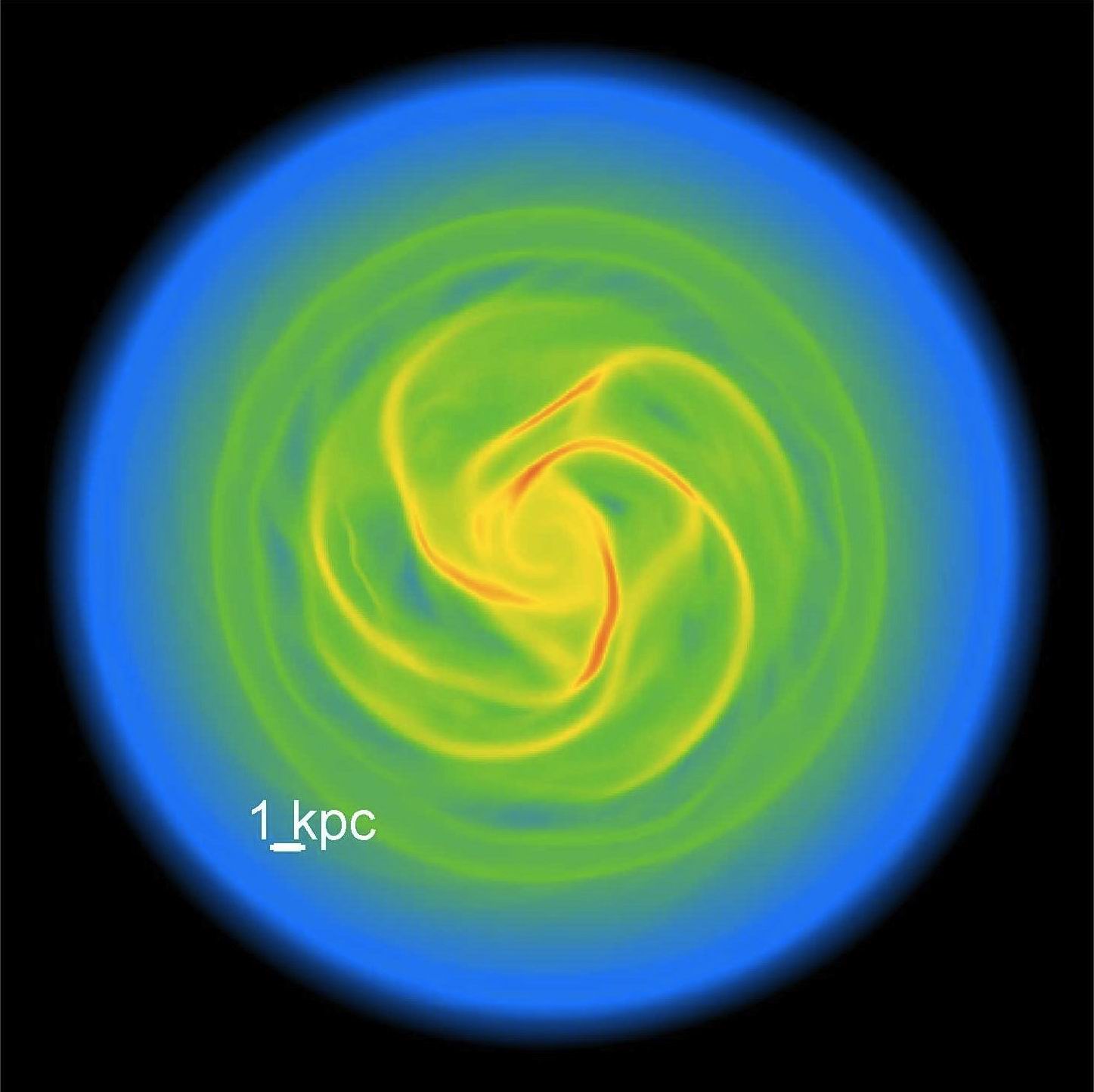}\includegraphics[width=0.25\hsize]{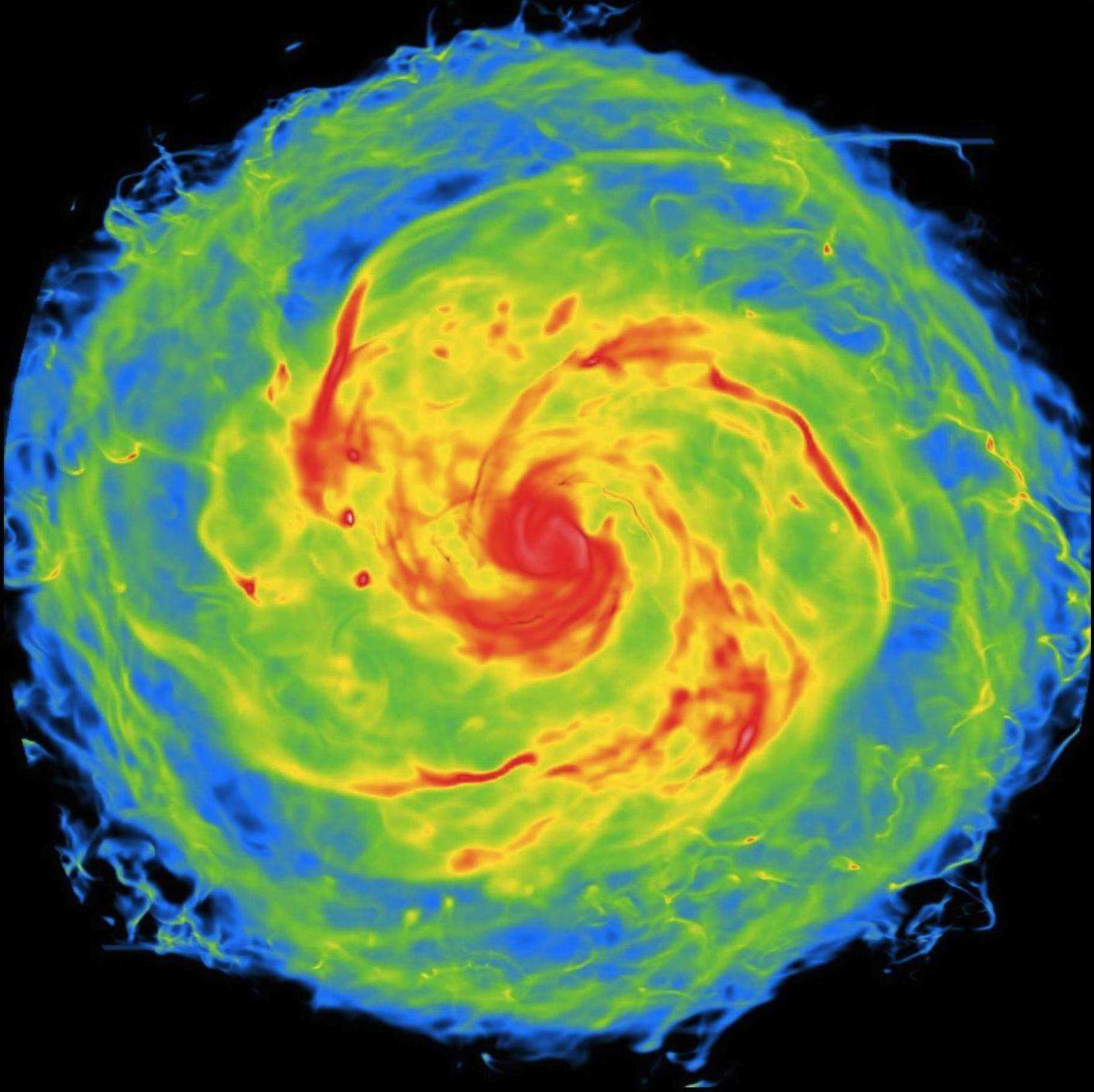}\includegraphics[width=0.275\hsize]{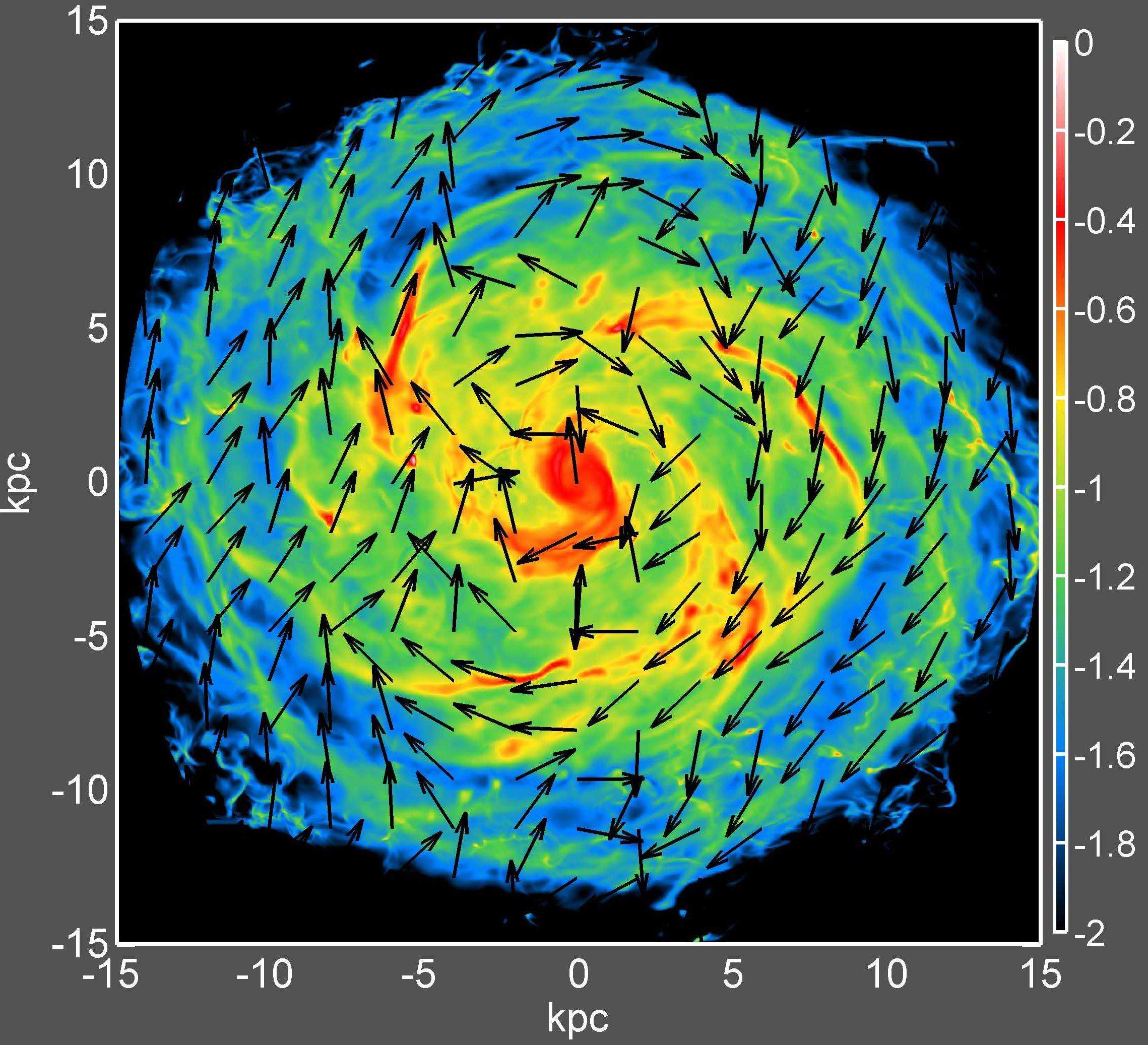}\\
\end{center}
\caption{The Galactic disc evolution. First three columns: stellar (top) and gaseous (bottom) distributions at $t=0, 300, 600$~Myr. Right pictures: radial variations of the module of magnetic field (top) and vectors of magnetic field in the disc plane.}\label{fig::galsim}
\end{figure}

Here we present an example of the gravitationally unstable stellar-gaseous galactic disc evolution. In this simulation 
we take into account magnetic field and radiation processes. The stellar disc is simulated using $N$-body method with 
number of particles $N=10^7$. The number of cells for gas dynamics equals to $1024\times 1024 \times 128$. We start 
our simulation from the equilibrium state of the galactic gaseous disc within the fixed potential of the dark matter 
halo for Toomre's parameter $Q_T = 1.2$. A construction of initial equilibrium configuration for stellar-gaseous 
disc is described by~\cite{2012MNRAS.427.1983K} in detail. We assume that the galactic disc consists of stellar component and gas with temperature $10^4$~K in the middle plane. The magnetic field is set as the toroidal 
configuration with $B = 6$~$\mu$G.
 
Figure~\ref{fig::galsim} presents the result of this simulation. The distribution of stellar particles (top row of 
three left panels) and gas surface density (bottom row of three left panels) are shown at time moments $t=0; 300; 600$~Myr.
One can see how the gravitational instability drives the formation of spiral pattern, which is clearly seen in both stellar 
and gaseous components. At $t=300$~Myr it is easily found narrow and dense galactic shocks (bottom row of panels). 
Because of relatively small initial value $Q_T$ a flocculent structure is rapidly developed in the spiral arms~\cite{2007A&A...473...31K}. The spatial
structure of a gas becomes very complex due to numerous shear flows, thermal and gravitational instabilities, 
magnetic field pressure influence as well. 

One can see the formation of the large number of dense gaseous complexes with number density $\sim 300-1000$~cm$^{-3}$. 
These complexes are mostly located in galactic spiral arms. This result is almost independent on numerical resolution 
and slightly dependent on the initial magnetic field. The increase of numerical resolution leads to formation of clouds
on smaller scales. Such clouds are expected to give birth of stars.

Figure~\ref{fig::galsim}(upper right panel) shows the structure of magnetic field at galactic disc at $t=600$~Myr. 
The amplitude of magnetic field decreases with radius from $|B| = 8 \mu G$ at $4$~kpc down to $|B| = 4 \mu G$ at the 
outer part of the disc. This is in a good agreement as with observations~\cite{2013MNRAS.433.1675B} and with cosmological
simulations~\cite{2013MNRAS.432..176P}. The vector map (right bottom panel in Figure~\ref{fig::galsim}) represents 
the chaotic and regular velocity components. The regular field corresponds to the rotation of a gas and large-scale 
spiral structure in the disc.

\section{Conclusion}

In this paper we have described our three-dimensional numerical code for multi-component simulation of the galactic evolution. 
This code has been mainly developed to study the evolution of disc galaxies taking into account generation of spiral
structure, physics of interstellar medium, formation of clouds and stars. Our code includes the following ingredients:
$N$-body dynamics, ideal magneto-gas dynamics, self gravity of gaseous and stellar components, cooling and
heating processes, star formation, chemical kinetics and multi-species gaseous and particle (for dust grains) advection. 
We present several tests for our code and show that our code have passed the tests with a resonable accuracy. 
It should be noted that our code is parallelized using the MPI library. We apply our code to study the large scale dynamics of galactic discs in context of formation and evolution of galactic spiral structure~\cite{2012MNRAS.427.1983K}, molecular clouds formation~\cite{2013MNRAS.428.2311K} and disc-to-halo interactions~\cite{2013MNRAS.431.1230K}.

\section{Acknowledgments}

The numerical simulations have been performed on the supercomputer 'Chebyshev' at the Research Computing Center~(Moscow State University) and MVS100k at the Joint Supercomputer Center~(Russian Academy of Sciences). This work was partially supported by the Russian Foundation of the Basic Research (grants 12-02-31452, 13-02-90767, 12-02-92704, 12-02-00685, 13-01-97062). SAK and EOV thanks to the Dmitry Zimin's "Dynasty" foundation for the financial support.

\section*{References}

\bibliography{ccpbib}

\end{document}